\documentclass[acmsmall,screen,nonacm]{acmart}

\usepackage{soul}
\usepackage{color}

\AtBeginDocument{%
  \providecommand\BibTeX{{%
    \normalfont B\kern-0.5em{\scshape i\kern-0.25em b}\kern-0.8em\TeX}}}

\usepackage[T1]{fontenc}
\usepackage{tikz}
\usetikzlibrary{arrows, calc, decorations.markings, positioning}
\input{timeline_pack}

\usepackage{adjustbox}
\usepackage{tabularx}
\usepackage{todonotes}

\usepackage{tcolorbox}

\usepackage{forest}
\usetikzlibrary{shadows}

\usepackage{color, colortbl}
\definecolor{Gray}{gray}{0.8}
\definecolor{LightGray}{gray}{0.95}

\newcounter{rowcntr}[table]
\renewcommand{\therowcntr}{\thetable.\arabic{rowcntr}}
\newcolumntype{N}{>{\refstepcounter{rowcntr}\therowcntr}c}

\begin{document}
\title{Machine Learning in Access Control: A Taxonomy and Survey}


\author{Mohammad Nur Nobi}
\affiliation{%
  \institution{Institute for Cyber Security (ICS) and Department of Computer Science,}
  \institution{University of Texas at San Antonio}
  \state{Texas}
  \country{USA.}
  }
\email{mohammadnur.nobi@utsa.edu}

\author{Maanak Gupta}
\affiliation{%
  \institution{Department of Computer Science,}
  \institution{Tennessee Tech University}
  \country{USA.}
}
\email{mgupta@tntech.edu}

\author{Lopamudra Praharaj}
\affiliation{%
  \institution{Department of Computer Science,}
  \institution{Tennessee Tech University}
  \country{USA.}
}
\email{lpraharaj42@tntech.edu}

\author{Mahmoud Abdelsalam}
\affiliation{%
  \institution{Dept. of Computer Science,}
  \institution{North Carolina A\&T State University}
  \country{USA.}
}
\email{mabdelsalam1@ncat.edu}

\author{Ram Krishnan}
\affiliation{%
  \institution{ICS, NSF Center for Security and Privacy Enhanced
Cloud Computing (C-SPECC), and Department of Electrical and Computer Engineering,}
  \institution{University of Texas at San Antonio}
  \state{Texas}
  \country{USA.}
  }
\email{ram.krishnan@utsa.edu}

\author{Ravi Sandhu}
\affiliation{%
  \institution{ICS, C-SPECC, and Dept. of Computer Science,}
  \institution{University of Texas at San Antonio}
  \state{Texas}
  \country{USA.}
  }
\email{ravi.sandhu@utsa.edu}

\renewcommand{\shortauthors}{Nobi et al.}

\begin{abstract}
An increasing body of work has recognized the importance of exploiting machine learning (ML) advancements to address the need for efficient automation in extracting access control attributes, policy mining, policy verification, access decisions, etc. 
In this work, we survey and summarize various ML approaches to solve different access control problems. We propose a novel taxonomy of the ML model's application in the access control domain. We highlight current limitations and open challenges such as lack of public real-world datasets, administration of ML-based access control systems, understanding a black-box ML model's decision, etc., and enumerate future research directions.
\end{abstract}



\maketitle

\section{Introduction}
\label{sec:introduction}

Machine Learning (ML) has shown to be extremely successful in solving tedious problems across many domains. Due to its innate ability to capture complex properties and extract related information, processes developed manually by a human can not even compete with an ML-based system. 
Like other domains, the access control field has many issues where a manual human-driven solution could achieve at best a sub-optimal solution. Consequently, there are huge avenues for improvement in different sub-areas of access control domain. Over time, access control researchers took the initiative to utilize the power of ML to obtain more efficient solutions. For example, several research works proposed ML-based solutions to design access control policies that are more robust than the traditional approaches~\cite{cotrini2018mining, karimi2021automatic, bui2020learning, frank2008class, bui2020decision, zhou2019automatic, jabal2020polisma}.
Besides, process automation is another important application of machine learning. Instead of using ML to solve a whole problem, ML can also assist in solving parts of the domain. Researchers exploited the advancements of ML to automate underlying laborious tasks in access control such as attributes extraction from the information or text that exists in plain natural language~\cite{alohaly2018deep, alohaly2019automated, alohaly2019towards, narouei2017towards, heaps2021access}, mapping between roles and permissions~\cite{zhou2019automatic, ni2009automating}, security rules extraction from access logs~\cite{cotrini2018mining, karimi2021automatic, mocanu2015towards}, or even extracting access control policies from the user stories~\cite{heaps2021access}.
Recent years have also witnessed the use of machine learning for access policy verification that could enhance the quality of implemented access control policy and ease the underlying burdens by automating the processes~\cite{heaps2019toward, hu2021verificationNIST}. It is also researched how machine learning-based method can automatically monitor the existing deployed access control policy and warns system administrators if it finds any suspicious activities~\cite{xiang2019towards}.
Moreover, researchers have proposed utilizing machine learning for access control decision making where a trained ML model will decide whether an access request is to be granted or denied~\cite{cappelletti2019quality, liu2021efficient, karimi2021adaptive, chang2006access, nobi2022toward}. 

As evident, the effect of ML in access control is found to be positive and has huge potential to achieve even more. Since the use of ML for access control purposes is rapidly emerging, there are several challenges encountered in the current ML oriented access control research practice. We observe that researchers applied ML methods on a case-by-case basis~\cite{frank2008class, karimi2018unsupervised, alohaly2019towards, cotrini2018mining, bui2019efficient, xiang2019towards}, thereby, there is no common strategy for using ML in the access control domain. The lack of an exhaustive overview of the application of ML in access control makes it even harder to gain in-depth insights into it and plan accordingly.
In addition, limited research efforts to determine the proper ML method for a distinct access control problem makes it challenging to select the best model for a novel problem.
Besides, there are several other limitations, such as a lack of quality datasets from the real-world organization~\cite{heaps2019toward, el2017abac} or the anonymous publicly available datasets may not contain relevant access control information capable of expressing a complete access control state of a system~\cite{molloy2011adversaries}. Putting all these into perspective, it is essential to have a holistic view of how researchers use machine learning for access control. Also, such a cohesive picture will help learn under-developed areas and determine future research in the domain.

In this paper, we perform a detailed review and summarize existing literature that uses ML to solve different access control problems, including attribute/roles or security policies extraction to access decisions and policy verification. To the best of our knowledge, this is the first comprehensive work which offers an encyclopedic view towards outlining the application of machine learning for access control system. We summarize our contributions to the field of access control as follows.
\begin{itemize}
    \item We perform a comprehensive review of existing access control literature that uses machine learning and discuss various research works done in different sub-domains of access control.
    \item We propose a novel taxonomy of machine learning in access control, and highlight research at each stage as the domain evolved chronologically.
    \item We summarize the publicly available real-world datasets used for machine learning based access control researches.
    \item We highlight open challenges and limitations faced by the research community as well as provide future research directions to thrive this critical security domain.
\end{itemize}

\subsection{Survey Methodology}
This survey is based on published research from 2006 to the first quarter of 2022. Before 2006, we did not find any ML-based access control solutions. We review papers from various search sources, including Google Scholar, ACM Digital Library, IEEEXplore, Springer, and preprints from arXiv. We use different keywords to filter related articles that use machine learning to solve \emph{any} access control related problem. 
This comprehensive review summarizes different approaches utilizing machine learning methods to solve access control problems.
To that end, we ask the following research questions while reviewing other works.
\begin{enumerate}
    \item What are the target access control models? In particular, is the proposed ML approach applicable for the access control domain as a whole or only suitable to any particular model?
    \item What are the ML methods that the respective approach uses?
    \item Why does the respective approach uses ML, and to what extent ML method contributed?
    \item What are the input and output of the ML model? Especially, can the trained machine learning model make access control decisions, or does the corresponding method only use an ML to improve or automate some sub-processes? 
    \item What kind of data was used for training the ML algorithm? What kind of data pre-processing is required for a related approach?
    \item How well an ML method performed towards solving the desired access control problem?
\end{enumerate}

\subsection{Organization of the Survey}

The paper is organized according to our proposed taxonomy which is presented in Section~\ref{sec:access-control-model}. This section provides a brief overview of access control. In Section~\ref{sec:ml-methods-overview}, we summarize ML methods that are widely used in the access control domain. Section~\ref{sec:categorization} demonstrates a timeline of how the application of ML evolved in access control context. In the same section, we review different approaches using machine learning to solve access control problems and briefly discuss each method following the proposed taxonomy. In Section~\ref{sec:open-challenges}, we review open challenges and future research directions in the context of access control and machine learning. Finally, Section~\ref{sec:conclusion} concludes our discussion.

\section{Overview of Access Control}
\label{sec:access-control-model}
\subsection{Access Control Preliminaries}


As the cornerstone of any information system's security, access control restricts the authorization of subjects/users to perform operations on system's resources/objects. Different abstraction, models and mechanisms have been proposed to govern the access and operations on resources to better represent the access control state of a system at various granularity levels with better maintainability. We briefly discuss some mainstream access control approaches and models in this section.

\subsubsection{Access Control List (ACL)} 
The ACL is the low-level representation of the access control state of a system. An ACL informs a system of a user's access privileges to system objects/resources (e.g., a file or a directory) \cite{sandhu1994access}. Each object has a security property that connects it to its access control list. The list has an entry for every user with access rights to the system.
However, as each user and resource is managed separately based on an identifier, ACLs make it very difficult to administer.

\subsubsection{Discretionary Access Control (DAC)}
It is an access control model where the \textit{Owner} or \textit{Security Administrator} of a system sets policies defining who can access the system resources ~\cite{sandhu1994access,6585778}. DAC is efficiently implemented using ACL or \emph{capabilities} \cite{thomas1993discretionary}. The security administrator defines a resource profile for each object and updates the access control list for the profile. In DAC, it is difficult to enforce the principle of least privilege and separation of duties~\cite{downs1985issues}.

\subsubsection{Mandatory Access Control (MAC)}
In the MAC model, the administrator makes the policy decision, which is based on the security labels of subjects (clearance) and objects (classification) \cite{sandhu1993lattice}. The user does not control, define or change access rights. A common example of MAC is in military security, where the data owner does not decide on who has a top-secret clearance, or change the classification from top-secret to secret ~\cite{hu2021verificationNIST}.
In MAC, configuration and maintenance of the system is a challenging task for the administrator, along with covert channel attacks.

\subsubsection{Role Based Access Control (RBAC)} In RBAC, the user’s access permissions will depend on what role he/she has in the organization~\cite{485845,sandhu1998role,gupta2017object}. An administrator designs the role defining what permissions a role should have and which users will be assigned what roles. For example, some users may be assigned to a role which allows them to read and write a file, whereas other user may be assigned a different role, which constrains them to read the file only.
A user can be assigned many roles, and at the same time, a role can be assigned to many users. Roles in RBAC offer easy administration of permissions, as role \textit{clusters} the abilities a user can hold. RBAC is useful for such scenarios because, with RBAC, there is no need to change the policies whenever an employee leaves the organization or changes jobs. The administrator can remove the employee from the role or add him to the new role. Depending on his job profile, a new employee can be assigned to his role promptly. RBAC also offers static and dynamic separation of duties, with the ability to define other constraints such as mutually exclusive roles, permissions, and cardinality constraints. However, RBAC is vulnerable to role explosion problem, which can lead to leakage of permissions as the numbers of users and roles in an organization expands.  


\subsubsection{Attribute Based Access Control (ABAC)}
In ABAC, a user's access to a resource is determined based on the attributes of the user, attributes of the resource, environmental conditions, and a set of pre-defined policies \cite{hu2015attribute,servos2017current,gupta2020secure}.
An ABAC policy defines the combinations of the user, resource, or environmental attributes and eventually these attributes are required to allow the user to operate on a resource.  
When an access request is intercepted by the ABAC system, the system matches the request with the underlying ABAC policies, where all conditions and restrictions are defined. The access request will get the desired access if the request satisfy the policy. 
For example, to restrict a department's resource to its `sales\_manager' only, the ABAC policy could limit the rule so that users with `job\_role' attribute has value `sales\_manager' can access the respective resource. As the access policy is designed with respect to rules, ABAC is one of the most generalized, scalable, and reliable access control models~\cite{gupta2018attribute,gupta2019dynamic}. However, there are challenges while implementing ABAC~\cite{brossard2017systematic} as it involves multiple laborious sub-processes such as \emph{attribute engineering}, \emph{attributes assignment} to the user and objects, \emph{policy engineering}, etc. Therefore, a skilled human administrator must design ABAC rules to satisfy overall security policy requirements.

\subsubsection{Relationship Based Access Control (ReBAC)}
Relationship-based access control defines access decisions based on the \emph{relationships} between subjects \cite{fong2011relationship, bui2020learning, bui2019efficient} or objects.
The most well-known examples of relationship-based access control are social networks. On Facebook, for example, the user gives view access to the photos or videos to friends or friends of friends. But friends of those friends cannot view the photos. Thus, ReBAC allows access when the user has a certain relationship with other entities in the system. 
ReBAC offers more than other access control models because it grants access based on multi-relationship between entities and takes decisions for certain entities, not entity types. For example, the user can access a specific photo from the directory, not the total directory.

\subsection{ML in Access Control Domain}

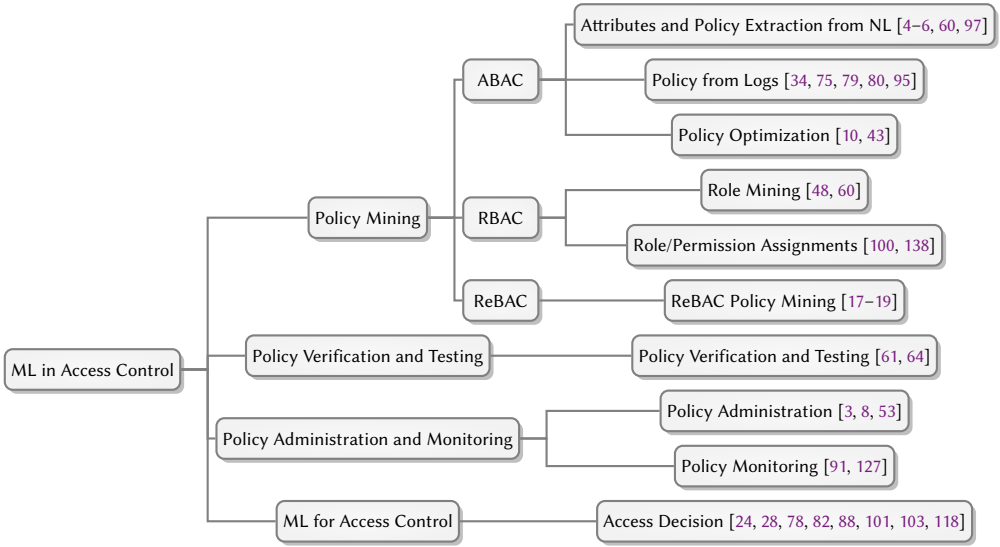
\begin{figure*}
    \scriptsize
    \vspace{-2ex}
    \centering
    \tikzset{
        my node/.style={
            draw=gray,
            inner color=gray!5,
            outer color=gray!10,
            thick,
            minimum width=1cm,
            rounded corners=3,
            text height=2.5ex,
            text depth=1ex,
            font=\sffamily,
            drop shadow,
        }
    }
\begin{forest}
    for tree={%
        my node,
        l sep+=5pt,
        grow'=east,
        edge={gray, thick},
        parent anchor=east,
        child anchor=west,
        if n children=0{tier=last}{},
        edge path={
            \noexpand\path [draw, \forestoption{edge}] (!u.parent anchor) -- +(10pt,0) |- (.child anchor)\forestoption{edge label};
        },
        if={isodd(n_children())}{
            for children={
                if={equal(n,(n_children("!u")+1)/2)}{calign with current}{}
            }
        }{}
    }
    [ML in Access Control
    [Policy Mining
    [ABAC[Attributes and Policy Extraction from NL~\cite{alohaly2018deep, alohaly2019automated, alohaly2019towards, narouei2017towards, heaps2021access}][Policy from Logs~\cite{mocanu2015towards, karimi2021automatic, karimi2018unsupervised, jabal2020polisma, cotrini2018mining}][Policy Optimization~\cite{benkaouz2016work, el2017abac}]]
    [RBAC[Role Mining~\cite{frank2008class, heaps2021access}][Role/Permission Assignments~\cite{zhou2019automatic, ni2009automating}]]
    [ReBAC[ReBAC Policy Mining~\cite{bui2019efficient, bui2020decision, bui2020learning}]]]
    [Policy Verification and Testing
    [Policy Verification and Testing~\cite{heaps2019toward, hu2021verificationNIST}]]
    [Policy Administration and Monitoring
    [Policy Administration~\cite{argento2018towards, alkhresheh2020adaptive, gumma2021pammela}][Policy Monitoring ~\cite{martin2006inferring, xiang2019towards}]]
    [ML for Access Control
    [Access Decision~\cite{chang2006access, cappelletti2019quality, khilar2019trust, liu2021efficient, outchakoucht2017dynamic, srivastava2020machine, karimi2021adaptive,nobi2022toward}]]
    ]
\end{forest}
    \caption{
    A Taxonomy of Machine Learning in Access Control Domain. NL indicates Natural Language.}
    \label{fig:taxonomy-access-control-pipeline}
    \vspace{-2ex}
\end{figure*}

\subsubsection{Policy Mining}
Policy Mining is the automation process of access control policy extraction from an existing access control state. Migration to a new model manually can be an error-prone and time-consuming task. To reduce the effort and cost of this task, extracting policies from the given access control information can be partly or completely automated~\cite{iyer2018mining}.
For example, if an organization had already an access control model implementation such as ACLs, and wants to migrate to more flexible, fine-grained access control model such as RBAC, ABAC, or ReBAC. In this case, an ABAC or ReBAC policy mining approach can be applied to obtain desired policies in automated and efficient manner~\cite{xu2014mining}. In general, the output of ABAC or ReBAC mining is a set of rules~\cite{bui2020learning}, whereas, the RBAC policy mining approaches produce permission to roles (PA) and user to roles (UA) assignments~\cite{zhou2019automatic, ni2009automating}.
There are several approaches that use advancements in machine learning to automate the ABAC~\cite{alohaly2019automated, alohaly2019towards, karimi2018unsupervised, jabal2020polisma, cotrini2018mining}, RBAC~\cite{frank2008class, heaps2021access, zhou2019automatic}, and ReBAC~\cite{bui2019efficient, bui2020decision, bui2020learning} policy mining process.


\subsubsection{Policy Verification and Testing}
Access control policy verification confirms no flaws within the policy that leaks or blocks access privileges. As a software test, access control policy verification depends on model proof, data structure, system simulation, and test oracle to validate that the policy logic functions are working properly. However, these methods have capability and performance concerns related to inaccuracy and complexity limited by applied technologies. For instance, model proof, test oracle, and data structure methods assume that the policy under verification is flawless unless the policy model cannot hold for test cases. Thus, the challenge of the method is to compose test cases that can systematically learn all faults. Moreover, a system simulation method needs to convert the policy to a simulated system. But, if the policy logic is complex or the number of policy rules is large, the translation between the system may be difficult or impractical \cite{hu2021verificationNIST}.
To overcome these challenges machine learning approach is used for policy verification which does not require comprehensive test cases, oracle, or system translation. Rather, it checks the logic of the policy rule directly, making it more efficient and feasible compared to the traditional method~\cite{heaps2019toward, hu2021verificationNIST}.

\subsubsection{Policy Administration and Monitoring}

Designing the policies is not an easy task and requires substantial administrative effort to modify the policy to accommodate changes. There are multiple ML-assisted methods to support additional adjustments in the existing access control policy and help detect improper behavior in the access control system.
ML based techniques can help system administrators to reduce the cost of generating the policies and adjust them dynamically~\cite{gumma2021pammela, alkhresheh2020adaptive}.
Besides, to accommodate changes in the system, the system admin may introduce errors and misconfiguration to the system due to the dependencies on user, data, functionality, and domain~\cite{argento2018towards, xiang2019towards}. Despite efforts on verification and testing~\cite{hu2021verificationNIST} to avoid access control misconfiguration, it is still difficult to avoid it in the real world system. An automated real-time monitoring system could help to observe the changes in access control behavior better and act accordingly~\cite{xiang2019towards, martin2006inferring}.

\subsubsection{ML for Access Control}
The massive policy scale and number of access control entities in open distributed information systems, such as Big Data, Internet of Things, and cloud computing, makes it impossible to list all possible rules and make access decision evaluating them. To ease the burden of implementing and maintaining access-control aspects in a system, there are multiple efforts to automate the access decision process using ML methods~\cite{liu2021efficient, karimi2021adaptive, cappelletti2019quality, nobi2022toward}. In all such cases, the system trains an ML model based on available logs, access history information, or existing access control state. Later on, the trained ML model decides accesses.

\section{Overview of Machine Learning}
\label{sec:ml-methods-overview}
ML is a sub-domain of Artificial Intelligence that automatically improves with data and experience study.
Supervised machine learning is a method in which we put the labeled data into an ML model~\cite{zhao2007spectral}. The model is trained with known data to predict future outputs appropriately. Before feeding to the machine learning model, we must prepare data to enhance its quality. There are various supervised machine learning models such as Decision Tree, Random Forest, Support Vector Machine, Logistic Regression, etc. It requires a lot of training time to prepare the data. Sometimes it is difficult to label the real-time data and fill the gaps.

Unsupervised learning is an ML method that uses neither classified nor labeled information~\cite{sathya2013comparison}. The job of the ML model is to group unsorted information corresponding to similarities, patterns, and differences without any earlier training of data. There are various types of unsupervised ML, such as K-means clustering, Principal Component Analysis, and Independent Component Analysis. The unsupervised ML is computationally complex and less accurate than supervised Methods.

This section briefly introduces the common ML methods widely used in access control-related research. Based on the structure of the ML models, we discuss them splitting into two classes--- (1) symbolic approaches: which represent the learned information with highly-structured knowledge, e.g., rules, trees, etc. (2) non-symbolic approaches: a highly complex function represents the underlying ML model. In the former case, the model and its underlying rules are easily intelligible due to the structure of data in human understandable form, hence, also referred to as the ``white-box'' model. In the latter case, the models are often referred to as ``black-box'' due to the inability to understand a logic of the function or the model. We further discuss them below.

\subsection{Symbolic ML Methods}
\label{sec:symbolic-ml}

\subsubsection{Decision Tree (DT)}
Decision trees are one of the powerful methods normally applied in several fields, such as machine learning, image processing, and identification of patterns \cite{charbuty2021classification, geurts2006extremely}. A decision tree is a tree-based method in which each path beginning from the root indicates a sequence of data splitting until a Boolean outcome is reached at the leaf node. Each path in the decision tree is a decision rule that reasons the underlying decision of input. In contrast, the entire tree corresponds to a compound Boolean expression which involves conjunctions and disjunctions to get Boolean decisions~\cite{yang2019extended}.
Though the decision tree is simple to comprehend, there are a few limitations such as lower accuracy, less generalization, etc.


\subsubsection{Random Forest (RF)}
RF is a supervised form of learning that can deal with classification and regression problems. It is an effective decision tree ensemble used for large-scale and multivariate pattern recognition~\cite{breiman2001random}. This collective learning is determined based on the idea of the random subspace method and the stochastic discrimination method of classification. 
Random Forests struggle with multi-valued attributes in several dimensions. One of the common issues in RF is over-fitting, mainly in regression tasks. 

\subsubsection{K-Nearest Neighbours (KNN)}
K-nearest neighbors (KNN) algorithm is a supervised ML algorithm that can be applied for both classification and regression predictive problems~\cite{djenouri2019adapted,xin2018machine}. K-nearest neighbors (KNN) algorithm utilizes `feature similarity’ to estimate the values of new data points, which shows that the new data point will be assigned a value based on how closely it matches the points in the training set. 
Though the KNN algorithm is easy to implement, it has some disadvantages. In a large dataset, the cost of calculating the distance between the two-point is huge, which diminishes the performance.

\subsubsection{K-modes and K-means}
K-Means clustering is an unsupervised learning algorithm. K-Means construct the clusters the objects that share similarities and dissimilar objects that belong to another cluster. The K-means algorithm represents the number of clusters represented by ``K'' in the K-means algorithm. In K-Means clustering, the data points are allocated to a cluster to minimize the sum of the square distance between the data points and the centroid. That is, a small variation within the clusters indicates more similar data points within the same cluster.
K-means algorithm requires prespecify the number of clusters. K means performing well for spherical shape. The algorithm is ineffective in identifying clusters with curving or complex geometric shapes. Because the K means algorithm does not allow the data points that are far from each other to share a similar cluster, even though they belong to one cluster~\cite{likas2003global}.

K-mode is an extension of k-means algorithm~\cite{huang1998extensions}. The k-means algorithm is
suitable for numerical data and does not perform well for categorical data because of the inadequate spatial representation. But the K-mode algorithm performs well in this situation \cite{huang1999fuzzy}.

\subsubsection{APRIORI-SD (Apriori Subgroup Discovery)}
The \emph{rule learning} algorithms~\cite{cohen1995fast} are designed to create classification and prediction rules. A \emph{subgroup} could be considered as most interesting if the group is as large as possible and
have the most unusual statistical (distributional) characteristics with respect
to a property of interest.
The APRIORI-SD ~\cite{kavvsek2006apriori} method adapts the rule learning to subgroup discovery~\cite{wrobel1997algorithm}. The APRIORI-SD finds population subgroups that are statistically \emph{most interesting}.
The algorithm works in three different phases to classify a set of samples. First, it computes a set of rules covering a certain number of samples in each rule. Then, the algorithm measures the \emph{confidence} for each rule and opt-out any rule if the measured confidence is below a pre-determined threshold. Finally, the algorithm determines a subset of filtered rules in the second phase based on the highest \emph{weighted relative accuracy}. Overall, the APRIORI-SD produces rules that are well generalized and smaller in size~\cite{cotrini2018mining, kavvsek2006apriori}.

\subsubsection{Gradient Boosting Method (GB)}
The Gradient Boosting technique is used for Regression and classification problems. This technique predicts in the form of a collection of weak prediction models \cite{chen2015xgboost}. Like any other supervised method, the GB method utilizes a feature vector to predict an output variable through a probability distribution. Thus a labeled data set is required, and the goal is to find an approximation function in the form of a weighted sum of functions from various classes of weak learners (the decision tree with fewer branches)~\cite{friedman2001greedy}. Though it is an important method to solve many classifications and regression problems, it is vulnerable to over-fitting issues.

\subsection{Non-symbolic ML Methods}
\label{sec:non-symbolic-ml}

\subsubsection{Support Vector Machine (SVM)}
SVM algorithm is one of the supervised machine learning algorithms supported by statistical learning theory~\cite{abdullah2021machine}. It chooses from the training samples a group of characteristic subsets, so the classification of the character subset is adequate for the division of the whole dataset~\cite{chang2011libsvm}. The SVM has been applied to resolve various classification problems effectively in many applications. Examples include intrusion detection, classification of face expression, prediction of your time series, speech recognition, image recognition, signal processing, detection of genes, text classification, recognition of fonts, diagnosis of faults, qualitative analysis, recognition of images, and other fields. However, for large-scale data processing in SVM, the time complexity and space complexity increase linearly with the rise in data. Thus, SVM is more capable of solving smaller samples, like nonlinearity and high dimensional problems.

\subsubsection{Restricted Boltzmann Machine (RBM)}
A restricted Boltzmann machine is a two-layer undirected graphical model where the first layer contains observed data variables (or visible units), and the second layer contains latent variables (or hidden units)~\cite{zhang2018overview}. The visible layer is fully connected to the hidden layer via pair-wise potentials, while the visible and hidden layers are confined to having no within-layer connections.
RBM is utilized for feature selection and extraction to solve many other tasks like dimensionality reduction, classification, and regression.

\subsubsection{Neural Network and Multi-Layer Perceptron (MLP)}
A neural network consists of neurons, input layers, hidden layers, and output layers \cite{carlini2017towards}. The neurons connect and have an associated weight and threshold. If the calculated result of any individual layer exceeds the threshold value, it can pass the value to the next layer. Otherwise, no data is passed to the next layer \cite{leofante2018automated}.

MLP is a neural network where the mapping between input and output is non-linear ~\cite{verma2020machine,hasan2020diabetes}. A multi-neuron perceptron consists of input and output layers and one or more hidden layers stacked together. MLP is useful for studying complex non-linear problems and large data input. However, the functional advantages of the model highly correlated with the quality of the model training.

\subsubsection{Convolutional Neural Network (CNN)}
In deep learning, a convolutional neural network (CNN/ConvNet) is a class of deep neural networks~\cite{o2015introduction}. In CNN, mostly the inputs are images, which allows us to determine specific properties in the architecture to distinguish certain patterns in the images. The CNNs utilize the spatial nature of the data to identify the objects, like the human perception of several objects in nature. A CNN consists of three layers: convolutional layers, pooling layers, and fully connected layers. When these layers are assembled, a CNN architecture has been formed.
\subsubsection{Recurrent Neural Network (RNN)}
RNN is a type of neural network that remember the data sequence and utilize data patterns for the prediction \cite{sherstinsky2020fundamentals}. 
RNN utilizes feedback loops that are different than other neural networks. With these loops, RNN process the sequence of the data. This loop collects the data to be shared from various nodes and give predictions. Recurrent neural networks are used to solve the machine learning problems such as speech recognition, language modeling, and image classification.

\subsubsection{Reinforcement Learning (RL)}
RL indicates a set of algorithms that train an agent to make a sequence of decisions through an interaction with an unknown environment to attain a goal (i.e., to increase the estimated collective discounted reward) \cite{karimi2021adaptive}. At each time, the agent examines the recent state. The agent takes action according to a policy. Following the action, the agent receives a reward signal. The agent's goal is to attain a policy that maximizes the expected reward, which is the sum of future discounted rewards. 
Though Reinforcement learning is a useful approach to solving many real-world problems, too much reinforcement can overload states which can diminish the results.

\subsubsection{Residual Neural Network (ResNet)}
One of the common challenges of deeper neural networks is the training difficulty. ResNet overcomes this challenge of training for substantially deeper networks by explicitly reformulating the layers~\cite{he2016deep}. It allows for training extremely deep neural networks with more than 150 layers.
A typical ResNet network is formed with stacks of multiple residual blocks followed by a Global Average Pooling layer. Each block is built on multiple convolutions whose output is added to the residual block's input and fed to the next layer. 
ResNet uses a skip connection to add the output from an earlier layer to a later layer.

\subsubsection{Bidirectional LSTM (BiLSTM)}
A BiLSTM is a structural processing model that contains two LSTMs: one getting the input in a forward direction and the other in a backward direction\cite{liu2016learning}. BiLSTMs efficiently enhance the quantity of information available to the network, increasing the context available to the algorithm \cite{huang2015bidirectional}.
Bi-LSTM is usually needed where the sequence-to-sequence tasks are required. A Bi-LSTM method is also effective in text classification, speech recognition, and forecasting models.

\subsubsection{Transformers}
A transformer is a process of transforming one sequence into another with the help of an encoder and decoder. BERT (Bidirectional Encoder Representation From Transformer), a notable NLP model, consists of several transformers encode layers. With more encode layers, it performs better in several language understanding task~\cite{devlin2018bert}.

\subsection{ML Evaluation Metrics}
It is the way to evaluate the performance of a classification problem where the output can be one of two possible classes, that is grant or deny. The access control element may have real and predicted results and both the elements have cases like True Positive (TP), True Negative (TN), False Positive (FP), False Negative (FN). From access control decision-making perspective, we can define these metrics as follows:
\begin{itemize}
    \item TP: The number of access requests that are correctly authorized by ML model/ policy.
    \item TN: The number of access requests that are correctly denied by ML model/ policy.
    \item FP: The number of access requests that are incorrectly authorized by ML model/ policy.
    \item FN: The number of access requests that are incorrectly denied by ML model/ policy.
\end{itemize}
\subsubsection{Accuracy}
The accuracy is defined as the number of correct predictions made by ML model. The accuracy can be calculated by the following formula:\[
Accuracy = \frac {TP+TN} {TP+FP+TN+FN}
\]
\subsubsection{Precision}
Precision represents the number of correctly permitted samples to the total number of predicted permitted samples. It is calculated by the formula:\[
Precision = \frac{TP} {TP+FP}
\]
\subsubsection{Recall or True Positive Rate (TPR)}
TPR is the number of correctly permitted sample to the total number of real permitted sample. Recall is calculated by the following formula:\[
TPR = \frac{TP} {TP+FN}
\]
\subsubsection{False Positive Rate (FPR)}
FPR is the number of wrongly permitted sample to the total number of wrongly permitted sample. FPR is calculated by the formula: \[
FPR = \frac{FP} {FP+TN}
\]
\subsubsection{F1 score}
F1 score is the weighted average of precision and Recall.
\[
F1=\frac {2\ast precision \ast Recall} {Precision+Recall}
\]

\noindent Policies (or models) with a higher F1 score lead to better generalization~\cite{cotrini2018mining}. Also, the higher TPR and Precision means how correctly and efficiently the policies (or models) can grant access. On the other hand, the policies (or models) with a lower FPR are better as they are less likely to give access to requests, which should be denied according to the ground truth policy.

\section{Machine Learning in Access Control}
\label{sec:categorization}
\begin{figure*}
\centering
\vspace{-2ex}
\input{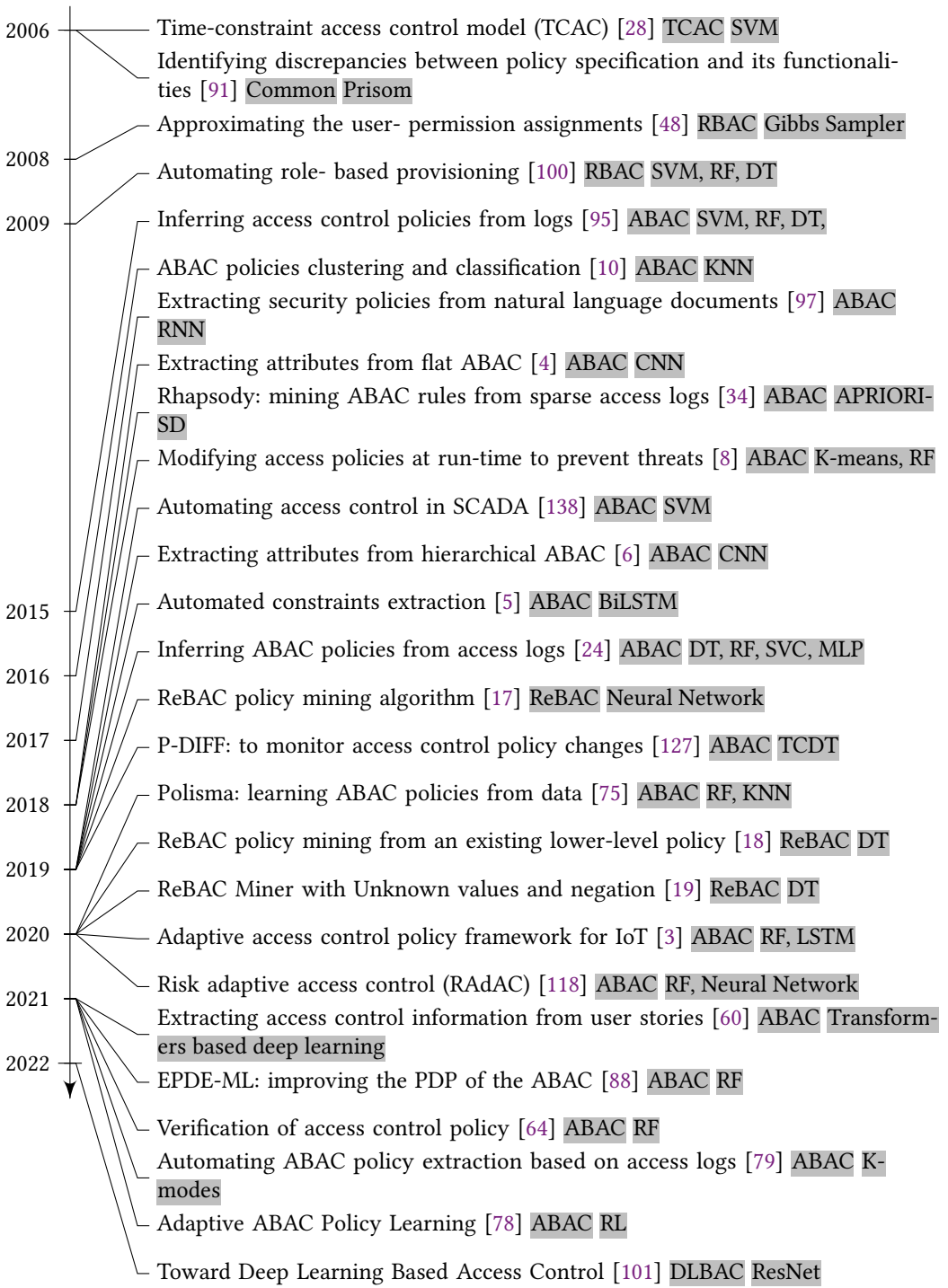}
\caption{
A Timeline of Seminal Works Towards ML in Access Control. In each work, the first grey highlight indicates the access control model (`Common' implies the method is applicable for any access control model), and the second highlight denotes ML algorithms applied in the corresponding method.
}
\label{fig:timeline-seminal-works}
\vspace{-2ex}
\end{figure*}

This section briefly describes literature using machine learning to solve access control-related problems. We organize the whole section according to the proposed taxonomy in Figure~\ref{fig:taxonomy-access-control-pipeline}. Before going into further details, we first present the timeline of seminal works using machine learning for access control in Figure~\ref{fig:timeline-seminal-works}. The timeline, sorted according to the published year, clearly illustrates how the application of machine learning evolves in the access control domain. As we see in Figure, machine learning in access control is emerging fast, and researchers published most of the work in recent years. Also, we summarize all the publicly available datasets in Table~\ref{tab:real-world-dataset}. The next section describes approaches developed for facilitating policy mining areas. The following sections discuss access control policy verification-related methods followed by policy enforcement-related works. 

\subsection{Policy Mining}
Due to the flexible policies and fewer management burdens, high-level access control models such as ABAC and ReBAC are adopted to support dynamic and complex security policies. However, adopting such policies from an existing lower-level policy like ACLs is challenging.
Generally, the policy mining techniques take attributes of users/resources in the system and the current access control state of the system as input. For ABAC and ReBAC, the algorithm output a set of rules (policy) that grants the same permissions~\cite{bui2019greedy, bui2017mining}. In contrast, RBAC mining algorithms output permission to roles (PA) and user to roles (UA) assignments~\cite{zhou2019automatic, ni2009automating}. Table~\ref{tab:ACPolicyMining} summarizes
access control policy mining approaches using machine learning.

\subsubsection{Attribute Based Access Control (ABAC)}

\textbf{\\Attributes and Policy Extraction from Natural Language.}
Natural language policies, being the preferred expression of policy~\cite{alohaly2018deep}, need to be transformed into a machine-readable form. Several researchers attempted to process such policies to extract access control-related information, including identifying policy sentences, triples of subject-object-action, etc. While manual extraction of such information is inefficient as the task becomes repetitive, requires more time, and is error-prone, several other approaches have been proposed to automate the process~\cite{narouei2017towards, narouei2015towards, slankas2013access, slankas2014relation, xiao2012automated}.

Narouei et al. \cite{narouei2017towards} introduce a top-down policy engineering framework for ABAC that automatically extracts security policies from natural language (NL) documents using deep recurrent neural networks (RNN). The paper uses high-level requirement specification documents referred to as natural language access control policies (NLACPs) which define statements governing management and access of critical objects in an enterprise. These NLACPs are simple human-readable expressions translated into machine enforceable access control policies. This work creates a deep recurrent neural network that uses pre-trained word embeddings to identify sentences that contain access control policy content. The authors develop a dataset from real-world policy documents from where a randomly selected subset of 2660 sentences is used and annotated for the presence of access control policy (ACP) content. The RNN models' input is word embeddings, where each `word' from the sentence is represented as a vector, and each of the `sentences' is represented as a vector of these word vectors. The output of the model is ACP sentences from large requirements documents. Further, the authors compare the proposed approach to other methods, outperforming the SVM model results and improving 5.58\% over the state-of-the-art. A prime limitation of this work is the involvement of the human factor in determining the correct identifiers of ACP sentences from NL documents. 

Alohaly et al.~\cite{alohaly2018deep} propose a deep learning-based automated approach for extracting ABAC attributes from NL policy. The authors develop a framework to automate the attribute extraction and related information leveraging natural language processing (NLP), relation extraction (RE), and CNN.
The authors first locate the modifiers of each policy element by analyzing the NL requirements. It extracts the attribute value and corresponding policy element, then identifies the attribute's category, short name, and data type.
Authors apply CNN to capture subject-attribute and object-attribute element-specific relations. They first generate candidate instances involving common patterns and feed them into the CNN classifier to determine whether a candidate instance holds a relation or not. They utilize two separate binary classifiers to determine subject and object attribute relation. Authors apply word embedding to encode words to real-valued vectors using GloVe pre-trained word embedding~\cite{pennington2014glove}. Finally, a sigmoid layer determines the binary classification.
The authors evaluate the framework based on a set of NLACPs by augmenting some existing datasets, including iTrust~\cite{meneely2012appendix}, IBM Course Management App~\cite{ibmcourse2004}, CyberChair~\cite{van2012cyberchair}, and Collected ACP~\cite{xiao2012automated}. As the policies in the existing datasets are more role-based focused and do not have enough attributes information, the method expands either subjects or object elements to inject some additional ABAC context.
The authors apply the F1 score as the evaluation metric. Based on their evaluation, the F1 score of the subject's attributes extraction is 0.96, while the performance is lowered by 0.05 for the object's attributes.

\begin{figure}[t]
    \vspace{-2ex}
	\includegraphics[width=.6\linewidth]{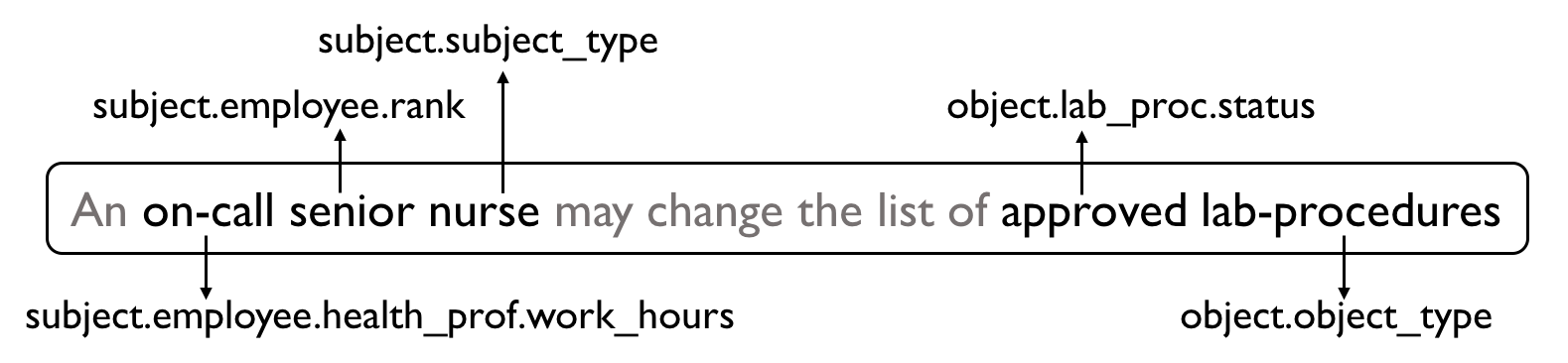}
	\caption{Attributes Identification from a Natural Language Hierarchical ABAC Policy.}
	\label{fig:hierarchical_abac_policy}
	\vspace{-2ex}
\end{figure}

The primary goal of the Alohaly et al.~\cite{alohaly2018deep} is to detect values of attributes (attribute extraction) in ABAC policies with no hierarchy among subject or object elements (\emph{flat ABAC}). From a practical perspective, extraction of authorization attributes of \emph{hierarchical ABAC} system from natural language artifacts is more needed than the flat counter-part.
The authors in~\cite{alohaly2019automated} extend the framework proposed in~\cite{alohaly2018deep}, built upon recent advancements in NLP and ML techniques, with automatic attributes extraction ability from NL hierarchical ABAC policies. 
The entire proposed framework works in multiple distinct phases. 

First, it extracts attribute value and related policy element. As shown in Figure~\ref{fig:hierarchical_abac_policy}, \emph{nurse} and \emph{senior} are the attribute values of two policy elements \emph{subject\_type} and \emph{rank}, respectively. 
Then, the algorithm extracts policy elements and attribute values capturing subject-attribute and object-attribute relations by identifying the grammatical patterns that encode each relation.
The next phase generates two lists of subject-attribute and object-attribute relations candidate instances utilizing commonly available patterns. The authors build two separate CNN-based classifiers for two relations based on respective candidate instances and determine whether a candidate instance encodes the corresponding relationship. The framework uses the GloVe pre-trained word embedding~\cite{pennington2014glove} to embed candidate instances.
Later on, the framework executes the second phase to suggest attribute's short names leveraging a density-based clustering method known as DBSCAN~\cite{ester1996density}. The authors utilize traditional algorithms to perform later steps that include hierarchically assigning the attributes to namespaces and identifying category and data type of attributes. 
For performance evaluation of the proposed technique, similar to Alohaly et al.~\cite{alohaly2018deep}, the authors construct a set of realistic synthetic NLACPs using some real-world datasets~\cite{meneely2012appendix, ibmcourse2004, van2012cyberchair, xiao2012automated}. Based on the experiments, the proposed framework shows a promising result with an average F1 score of 0.96 when extracting the subject's attribute values and 0.91 for object attribute values extracting.

In ABAC, the attributes often need to satisfy some constraints, which is critical for complying with organizational security policy. It requires extraordinary skills and efforts to define the constraints formally. The process is tedious and error-prone as security architects have to analyze several documents to develop them manually. 
The Alohaly et al.~\cite{alohaly2019towards} proposed a reliable and automated constraints extraction process exploiting tools in NLP. This automation also enables trace-ability formal constraints expressions and related policies that will also help to avoid errors while assigning unauthorized attributes.
First, the proposed method extracts the \emph{conflicting} factors, as the constraints are defined using the notion of \emph{conflict}~\cite{bijon2013constraints, bijon2013towards}. To determine the conflicts in NLACPs, the authors initially annotate the NLACPs by specifying different levels of conflict ranging from 0 to 3, which is expected to be done manually by security architects for each policy sentence. In the next phase, the method identifies the constraints expressions by determining the right \emph{boundaries} (inside, beginning, or outside) of conflicting values within the policy sentence. 

To this end, the authors apply the state-of-the-art Bidirectional Long-Short Term Memory (BiLSTM) model of Recurrent Neural Networks (RNNs)~\cite{huang2015bidirectional}. As input, the BiLSTM takes each word's real-valued vector representation in the NLACP sentence using GloVe pre-trained word embedding. Internally, the BiLSTM layer trains two distinct LSTMs. One is trained to acquire the sequential information of given input, while the other learns information encoded in the reverse order~\cite{alohaly2019towards}. Finally, the model assigns a \emph{label} for each word indicating its boundary, whether the word is in inside, at the beginning, or in the outside. The extraction process ends with the normalization of the constraints.
To evaluate the effectiveness of the proposed approach, the authors created a dataset (not publicly available) from real-world policy documents with authorization rules for various departments, including Human Resources, Information Technology, etc., from educational institutions. The dataset comprises 801 occurrences of constraints in 747 NLACP sentences, where each of the sentences contains at least one constraint. The experimental results imply that the proposed approach holds promise for enabling constraint automation with an F1 score of 0.91 in detecting at least 75\% of each constraint expression.

\begin{figure}[t]
    \vspace{-2ex}
	\includegraphics[width=0.7\linewidth]{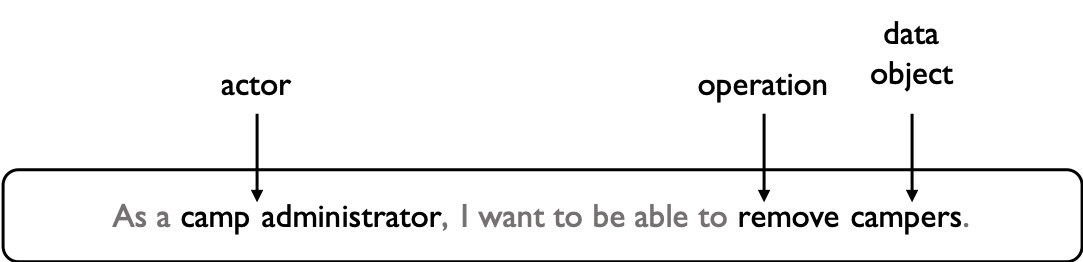}
	\caption{A sample user story.}
	\label{fig:user_story}
	\vspace{-2ex}
\end{figure}

The user stories written by software developers better represent the actual code than high-level product documentation. Such stories can contain access control-related information, and extracting that information can be used to construct access control documentation. Also, stakeholders can use this information for access control engineering, development, and review. For example, as shown in Figure~\ref{fig:user_story}, the user story contains access control information where `camp administrator' is the actor and `campers' is a data object that the camp administrator should be able to perform a `remove' operation. Authors in~\cite{heaps2021access} developed an automated process using ML to extract access control information from a set of user stories that present the behavior of the software product in question. The proposed work takes a collection of user stories as input to a transformers-based deep learning model~\cite{devlin2018bert} with three components: access control classification, named entity recognition, and access type classification. First, it classifies whether a user story contains access control information or not. Then, it identifies the actors, data objects, and operations in the user story as part of a named entity recognition task.
Eventually, the proposed technique determines the type of access between the identified actors, data objects, and operations through a classification prediction. The authors apply three distinct transformers-based learning models for the above three tasks. 

The authors experiment and evaluate the proposed method with Dalpiaz dataset~\cite{dalpiaz2018requirements,dalpiaz2018pinpointing} consisting of 21 web applications from various domains, each with 50–130 user stories for a total of over 1600 user stories. SL.~\ref{row:dalpiaz} of Table~\ref{tab:real-world-dataset} outlines details of this dataset. The authors compare their results with CNN and SVM-based implementation for all three tasks. The overall results suggest that the transformers perform the best in all categories. However, CNN performs as good as Transformers for the named entity recognition task. Though the Transformers model performs better than the CNN and SVM, the difference is not that notable. The authors suggest that the first step would be to have a larger and more robust dataset for performance enhancement.

\noindent
\textbf{Policy from Access Logs.}
Work by Mocanu et al.~\cite{mocanu2015towards} proposed a deep learning-based approach to infer policies from logs, which also supports negative authorization (i.e., a subject cannot access a resource). The proposed approach has two phases. The first phase generalizes the knowledge from logs providing a set of candidates rules in a binary vector format. In the second phase, this set of candidate rules is transformed to the format acceptable by Xu-Stoller~\cite{xu2014mining} and compared. The work uses Restricted Boltzmann Machines (RBMs) as density estimators to propose a technique capable of producing a set of suitable candidate rules based on the knowledge extracted by the RBMs from the processed logs. The authors use a healthcare dataset to evaluate the effectiveness of their approach. A significant advantage of their approach is the support for negative authorization. Still, their proposed work is limited to the first phase, and more implementation results and evaluations are required for real-world policies containing various types of expressions.

However, such kinds of methods~\cite{xu2014mining, mocanu2015towards} are limited where the logs have only a tiny subset of all possible access requests. Evidently, the access logs in real-world systems only contain about 10\% of all possible access requests~\cite{cotrini2018mining}. Over time, the developed rules also become highly complex due to its adoptions to different changes in the system. Hence, the shorter and less-convoluted rules are better from a maintainability perspective.
Cotrini et al.~\cite{cotrini2018mining} proposed an approach, known as Rhapsody, for mining ABAC rules from sparse access logs identifying patterns among the authorized requests. Rhapsody also solves various other issues of the existing state-of-the-art, including the rules size and over-permissiveness nature.
The Rhapsody builds upon APRIORI-SD~\cite{kavvsek2006apriori},which the authors modify to overcome its weaknesses of mining overly permissive or unnecessarily large rules. Rhapsody takes an ABAC instance as input and provides a set of rules as the output. It performs the overall computation in three different stages. At first, it determines a set of candidate rules that cover a certain amount of requests available in the logs. Next, it filters out overly permissive rules. The algorithm opt-out any rule from the rules set if there is an equivalent shorter rule.

The authors evaluate the performance of Rhapsody using two Amazon access logs~\cite{AmazonKaggle2013, AmazonUCI2011} and a log of students accessing a computer lab at ETH Zurich. Both organizations' logs are sparse where Amazons' logs show about 7-10\% of all the employees have requested access, whereas the computer lab logs have access of less than 1\% of the users. Besides, the authors also experiment with some synthetic datasets that contain approximately 50\% of all possible requests. 
Cotrini et al. also introduce a novel validation method named \emph{universal cross-validation} that evaluates mined rules with requests not present in the dataset. Experiments show that the proposed evaluation approach validates policies with higher F1 scores (more accurate in deciding requests outside the log) than the standard cross-validation technique. The authors compare the performance of Rhapsody with Classification Tree~\cite{breiman2017classification}, CN2~\cite{clark1989cn2}, and two state-of-the-art ABAC policy mining algorithms (Xu and Stoller~\cite{xu2014mining} and APRIORI-SD~\cite{kavvsek2006apriori}). The results signify that Rhapsody can mine concise rules from sparse logs with higher accuracy than state-of-the-art mining algorithms. It also ensures better generalization and avoids mining overly permissive rules.

Jabal et al.~\cite{jabal2020polisma} propose a novel framework for learning ABAC policies, named Polisma, combining data mining, statistical, and ML techniques. The methodology of the Polisma approach works mainly in four different stages. In the first step, a data mining technique generates a set of rules inferring associations between the users and resources. Next, the generated rules are generalized based on statistically significant attributes and context information. In the third step, the generalized rules are further augmented with `restriction rules' for limiting users' access to resources concerning the respective authorization domain. Policies learned by those three stages are `safe generalizations' (do not have unintended consequences) with limited \emph{over-fitting}. However, there might be other access requests for which the generated policy set cannot make an access decision. To overcome this issue, Polisma applies Random Forest (RF) and KNN as the machine learning classifiers on requests not covered by the learned set of policies and uses the classification result to label these data and generate additional rules. The approach is evaluated empirically using both the real-world~\cite{AmazonUCI2011} and synthetic datasets. Experimental results show that Polisma can develop ABAC policies that accurately control access requests.

Many large-scale businesses need to grant authorizations to their users that are \emph{distributed} across heterogeneous computing environments. In such a case, the manual development of a single access control policy framework for an entire organization is cumbersome and error-prone. Karimi et al.~\cite{karimi2021automatic} proposed an automating ABAC policy extraction based on access logs following their other work on an unsupervised learning-based approach for ABAC policy mining~\cite{karimi2018unsupervised}. The proposed method extracts policy rules containing positive and negative attributes and relationship filters. The negative filters are missing in existing policy mining approaches, which are essential in scenarios when an exception is expressed. The authors apply the K-modes algorithm~\cite{cao2009new} to divide the access log into clusters based on finding similar patterns between features (i.e., attribute values) of
the records (i.e., access control tuples). The records in each cluster correspond to one access control rule in the system. Next, the authors extend rules extraction algorithms proposed in~\cite{karimi2018unsupervised} to extract both positive and negative attribute filters. Rules extracted at this stage might not be accurate and concise desirably. To enhance the quality and accuracy of mined policy, the authors apply multiple iterations of the rule pruning and policy refinement process.

The empirical study based on both real-world and synthetic datasets demonstrates the ability of the proposed approach to mine high-quality policies with minimal impact on the size of
attributes and attribute values. The proposed algorithm also shows effectiveness with an incomplete set of access logs and handles a reasonable amount of noise applicable in real-world applications. However, the proposed method is based on K-modes clustering, where finding the correct number of clusters (K) is challenging. Also, the technique has dependencies on tuning multiple parameters that equally affect the performance. As a result, the authors do not claim that the proposed method will extract the policy with the highest quality in every scenario. However, by trial and error, more randomization in cluster initialization and a more comprehensive range of parameters will help to achieve optimal performance.

\noindent
\textbf{Policy Optimization.}
Benkaouz et al.~\cite{benkaouz2016work} presents an approach for ABAC policies clustering and classification. The proposed approach uses KNN algorithms that help to reduce dimensionality and achieve higher flexibility for ABAC policies in high-scale systems. The core idea is to create a cluster of similar policies given a set of ABAC policies. The value of the parameter \textit{k} (i.e., as defined for KNN algorithms, \textit{k} represents the number of nearest neighbors) is determined based on the criticality level of the application. This adds more flexibility to the ABAC model since the granularity of the ABAC policies could be adjusted based on the value of the parameter \textit{k}. A small \textit{k} implies fine-grained ABAC model, and a bigger \textit{k} implies a coarse-grained ABAC model. This is a work in progress, and several key questions remain unanswered, such as the default value of \textit{k}, the most suitable KNN algorithms for policies clustering, and also if the same approach has any uses in different kinds of applications. 

El Hadj et al.~\cite{el2017abac} propose ABAC-PC (ABAC Policy Clustering) that groups the policy rules according to the decision effects, such as to permit or deny rules. The method also creates cluster rules based on similarity scores and produces the minimum set of rules representing each cluster. The proposed work is an extension of the Benkaouz et al.~\cite{benkaouz2016work} method. The authors evaluate the proposed work suggesting that the reduction rate of policy rules can reach up to 10\% for an ABAC policy with over 9000 rules. In addition, the authors claim that the approach can be extended with other policy analysis tools to detect and resolve anomalies among XACML policies.

\subsubsection{Role Based Access Control (RBAC)}
\hfill\\

\textbf{Role Mining.}
The work in \cite{frank2008class} focuses on bottom-up approaches for RBAC role mining where the goal is to approximate the user-permission assignments by finding a minimal set of roles, user-role and role-permission assignments. Combinatorial algorithms can be used by identifying the differences from the original user-permission assignments but this often causes (1) presence of errors in the user-permission assignments and (2) identification of unmeaningful roles (e.g., roles should be meaningful to represent a particular job function within an organization). Instead, the authors propose a probabilistic framework to overcome the drawbacks of combinatorial approaches. The proposed framework works by generalizing the observations in the existing user-permission assignments. The authors use Gibbs sampler~\cite{neal2000markov} which is a Markov chain Monte Carlo algorithm for their proposed Disjoint Decomposition Model.

The authors use synthetic and real-world datasets for evaluation and compare the performance with a combinatorial algorithm. The synthetic dataset includes 200 users, 200 permissions, ten business roles, and five technical roles, and the user-permission assignments are randomly generated. The real-world (not publicly available) dataset includes 5000 users and 1323 permissions. Since there is no way to know the errors present in the real-world dataset, the authors apply a metric of role meaningfulness to compare models. The result demonstrates that the proposed approach creates meaningful roles and identifies erroneous user-permission assignments in given data.
One limitation of this work is that the authors uniformly introduce user-permission errors in the synthetic dataset, whereas the number of errors in real-world scenarios is unknown.

\AtBeginEnvironment{tabular}{\setcounter{rowcntr}{0}}

\begin{table*}[t]
\centering
  \caption{Publicly Available Real-world Datasets Used in Access Control Researches. The names presented in the `Name' column are chosen randomly.}
  \label{tab:real-world-dataset}
  \resizebox{\textwidth}{!}{%
  \begin{tabular}{Ncccccc}
    \hline
    \rowcolor{Gray}
    \multicolumn{1}{c}{SL.}
    & \multicolumn{1}{p{2cm}}{\centering Name}
    & \multicolumn{1}{p{1cm}}{\centering Publish Year}
    & \multicolumn{1}{p{3cm}}{\centering Reference}
    & \multicolumn{1}{p{2cm}}{\centering Type}
    & \multicolumn{1}{p{4.5cm}}{\centering Description}
    & \multicolumn{1}{p{2.2cm}}{\centering Application}\\
    \hline
    \label{row:ibm} & \multicolumn{1}{p{2cm}}{\centering IBM-CM}
    & \multicolumn{1}{p{1cm}}{\centering 2004}
    & \multicolumn{1}{p{3cm}}{\centering IBM~\cite{ibmcourse2004}}
    & \multicolumn{1}{p{2cm}}{\centering Access Policies}
    & \multicolumn{1}{p{4.5cm}}{\centering Natural language access control policy}
    & \multicolumn{1}{p{2.2cm}}{\centering \cite{alohaly2018deep}, \cite{alohaly2019automated}}\\
    
    \rowcolor{LightGray}
    \label{row:university-data} & \multicolumn{1}{p{2cm}}{\centering UniversityData}
    & \multicolumn{1}{p{1cm}}{\centering 2005}
    & \multicolumn{1}{p{3cm}}{\centering Fisler et al.~\cite{fisler2005verification}}
    & \multicolumn{1}{p{2cm}}{\centering Access Policy}
    & \multicolumn{1}{p{4.5cm}}{\centering Central grades repository system for a university}
    & \multicolumn{1}{p{2.2cm}}{\centering \cite{martin2006inferring}}\\
    
    \label{row:wikipedia} & \multicolumn{1}{p{2cm}}{\centering Wikipedia}
    & \multicolumn{1}{p{1cm}}{\centering 2009}
    & \multicolumn{1}{p{3cm}}{\centering Urdaneta et al.~\cite{urdaneta2009wikipedia}}
    & \multicolumn{1}{p{2cm}}{\centering Access Logs}
    & \multicolumn{1}{p{4.5cm}}{\centering Access request traces from Wikipedia}
    & \multicolumn{1}{p{2.2cm}}{\centering \cite{xiang2019towards}}\\
    
    \rowcolor{LightGray}
    \label{row:amazon-uci} & \multicolumn{1}{p{2cm}}{\centering AmazonUCI}
    & \multicolumn{1}{p{1cm}}{\centering 2011}
    & \multicolumn{1}{p{3cm}}{\centering UCI Repository~\cite{AmazonUCI2011}}
    & \multicolumn{1}{p{2cm}}{\centering Access Logs}
    & \multicolumn{1}{p{4.5cm}}{\centering Access data of Amazon employees}
    & \multicolumn{1}{p{2.2cm}}{\centering \cite{cappelletti2019quality}, \cite{cotrini2018mining}, \cite{jabal2020polisma}, \cite{karimi2021automatic}, \cite{nobi2022toward}}\\
    
    \label{row:iTrust} & \multicolumn{1}{p{2cm}}{\centering iTrust}
    & \multicolumn{1}{p{1cm}}{\centering 2012}
    & \multicolumn{1}{p{3cm}}{\centering Meneely et al.~\cite{meneely2012appendix}}
    & \multicolumn{1}{p{2cm}}{\centering Access Policies}
    & \multicolumn{1}{p{4.5cm}}{\centering Natural language access control policy}
    & \multicolumn{1}{p{2cm}}{\centering \cite{alohaly2018deep}, \cite{alohaly2019automated}}\\
    
    \rowcolor{LightGray}
    \label{row:cyber-chair} & \multicolumn{1}{p{2cm}}{\centering CyberChair}
    & \multicolumn{1}{p{1cm}}{\centering 2012}
    & \multicolumn{1}{p{3cm}}{\centering Stadt et al.~\cite{van2012cyberchair}}
    & \multicolumn{1}{p{2cm}}{\centering Access Policies}
    & \multicolumn{1}{p{4.5cm}}{\centering Natural language access control policy}
    & \multicolumn{1}{p{2.2cm}}{\centering \cite{alohaly2018deep}, \cite{alohaly2019automated}}\\
    
    \label{row:coll-acp} & \multicolumn{1}{p{2cm}}{\centering CollectedACP}
    & \multicolumn{1}{p{1cm}}{\centering 2012}
    & \multicolumn{1}{p{3cm}}{\centering Xiao et al.~\cite{xiao2012automated}}
    & \multicolumn{1}{p{2cm}}{\centering Access Policies}
    & \multicolumn{1}{p{4.5cm}}{\centering Natural language access control policy collected from multiple sources}
    & \multicolumn{1}{p{2.2cm}}{\centering \cite{alohaly2018deep}, \cite{alohaly2019automated}}\\
    
    \rowcolor{LightGray}
    \label{row:amazon-kaggle} & \multicolumn{1}{p{2cm}}{\centering AmazonKaggle}
    & \multicolumn{1}{p{1cm}}{\centering 2013}
    & \multicolumn{1}{p{3cm}}{\centering Kaggle~\cite{AmazonKaggle2013}}
    & \multicolumn{1}{p{2cm}}{\centering Access Logs}
    & \multicolumn{1}{p{4.5cm}}{\centering Two years historical access data of Amazon employees (12000 users and 7000 resources)}
    & \multicolumn{1}{p{2.2cm}}{\centering \cite{cappelletti2019quality}, \cite{cotrini2018mining}, \cite{karimi2021adaptive}, \cite{karimi2021automatic}, \cite{liu2021efficient}, \cite{nobi2022toward}}\\
    
    \label{row:e-document} & \multicolumn{1}{p{2cm}}{\centering eDocument}
    & \multicolumn{1}{p{1cm}}{\centering 2014}
    & \multicolumn{1}{p{3cm}}{\centering Decat et al.~\cite{decat2014document}}
    & \multicolumn{1}{p{2cm}}{\centering Access Policy}
    & \multicolumn{1}{p{4.5cm}}{\centering e-document case study}
    & \multicolumn{1}{p{2.2cm}}{\centering \cite{bui2020decision}, \cite{bui2020learning}, \cite{bui2019efficient}}\\
    
    \rowcolor{LightGray}
    \label{row:work-force} & \multicolumn{1}{p{2cm}}{\centering Workforce}
    & \multicolumn{1}{p{1cm}}{\centering 2014}
    & \multicolumn{1}{p{3cm}}{\centering Decat et al.~\cite{decat2014workforce}}
    & \multicolumn{1}{p{2cm}}{\centering Access Policy}
    & \multicolumn{1}{p{4.5cm}}{\centering Workforce management case study}
    & \multicolumn{1}{p{2.2cm}}{\centering \cite{bui2020decision}, \cite{bui2020learning}, \cite{bui2019efficient} }\\
    
    \label{row:scada-intrusion} & \multicolumn{1}{p{2cm}}{\centering SCADA-Intrusion}
    & \multicolumn{1}{p{1cm}}{\centering 2015}
    & \multicolumn{1}{p{3cm}}{\centering Turnipseed et al.~\cite{turnipseed2015new}}
    & \multicolumn{1}{p{2cm}}{\centering SCADA Data}
    & \multicolumn{1}{p{4.5cm}}{\centering SCADA dataset for intrusion detection system}
    & \multicolumn{1}{p{2.2cm}}{\centering \cite{zhou2019automatic}}\\
    
    \rowcolor{LightGray}
    \label{row:dalpiaz} & \multicolumn{1}{p{2cm}}{\centering Dalpiaz-UserStories}
    & \multicolumn{1}{p{1cm}}{\centering 2018}
    & \multicolumn{1}{p{3cm}}{\centering Dalpiaz et al.~\cite{dalpiaz2018requirements,dalpiaz2018pinpointing}}
    & \multicolumn{1}{p{2cm}}{\centering User Stories}
    & \multicolumn{1}{p{4.5cm}}{\centering Over 1600 user stories from 21 web applications}
    & \multicolumn{1}{p{2.2cm}}{\centering \cite{heaps2021access}}\\
    
    \label{row:incident} & \multicolumn{1}{p{2cm}}{\centering Incident}
    & \multicolumn{1}{p{1cm}}{\centering 2018}
    & \multicolumn{1}{p{3cm}}{\centering Amaral et al.~\cite{amaral2018enhancing}}
    & \multicolumn{1}{p{2cm}}{\centering Event Logs}
    & \multicolumn{1}{p{4.5cm}}{\centering Event log from an incident management process}
    & \multicolumn{1}{p{2.2cm}}{\centering \cite{cappelletti2019quality}}\\
    
    \hline
  \end{tabular}
  }
\end{table*}

\textbf{Role/Permission Assignments.}
The authors in~\cite{ni2009automating} propose a machine learning-based automated role maintenance system that provisions existing roles with entitlements from newly deployed applications and provisions new users with existing roles. The proposed technique utilizes the attributes of entitlements to simplify the management of entitlements assignments to roles. The authors solve the problem in the context of machine learning by defining a \emph{function} that maps between the various attributes of the entitlements and business roles. The goal is to learn the function to automatically provision business roles with entitlements by running a classifier based on that function. 
The authors accomplish the overall provisioning process in four different phases. First, it is required to collect existing role-entitlement mappings data from human experts. The collected data needs to be filtered for selecting essential attributes (features). It impacts the performance as all the features are irrelevant to the provisioning mapping. The filtered entitlements and users can belong to several roles (\textit{many-to-many}), or there might have some missing values. Hence, as the next step, it needs to pre-process the provisioning data to replace all missing values with a new value \textit{not applicable} and transform each \textit{many-to-many} mapping into several \textit{many-to-one} mappings. The classifier's training and selection of the best classifiers are performed in the third phase. The authors train multiple classifiers and compare their performance based on standard estimation methods (e.g., 10-fold cross-validation) to select the fittest classifiers for the next step. The provisioning process automatically classifies the given data to assign entitlements to roles in the final phase. 

To justify the efficacy of the proposed technique, the authors used both provisioning data (with 35 business roles, 659 mappings between the UI entitlements and the business roles) from a real-world organization and synthetic provisioning data. The real-world dataset is not available publicly. The authors experimented with 18 candidate classification algorithms, including support vector machines (SVM), RF, Decision Tree, etc., before choosing SVM as the final classifier. The authors used multiple evaluation metrics such as accuracy, false positive, and false negative. The performance results of SVM for a majority of the roles demonstrate that the FP rates are between 0\-5\%, and most FN rates are between 0 to 30\%. Also, 70\% or more of the assignments can be executed automatically, and no more than 30\% of the assignments will need some assistance.

Lu Zhou et al.~\cite{zhou2019automatic} propose two machine learning-based approaches to automate the role assignment process in the context of the real-time Supervisory Control and Data Acquisition (SCADA) system. The authors first demonstrate how SVM can be applied by developing a context-aware RBAC with automated role engineering. They use different static (job function, job position, etc.) and dynamic attributes (time, location, etc.) of the SCADA system’s users and devices as input to the SVM and obtain user-role or permission-role assignments as the output. The authors did not explain any detailed evaluation results for the SVM-based approach. However, the authors further experimented Adaboost algorithm for the role assignment automation process. Similar to SVM experimentation, the author used the same input (user and device attributes) to the Adaboost approach to obtain assigned roles or permission for the corresponding user. The authors use a dataset~\cite{turnipseed2015new} designed for intrusion detection in a SCADA system. The authors compared the performance of both the real and discrete-valued Adaboost algorithms in terms of accuracy, where they demonstrated results varied significantly for different experiments.

\subsubsection{Relationship Based Access Control (ReBAC)}
Like ABAC policy mining approaches, ReBAC policy mining algorithms can also potentially reduce the effort to obtain a high-level policy from lower-level access control data. 
The author in~\cite{bui2019efficient} propose an efficient and scalable ReBAC policy mining algorithm which is an extension over the existing evolutionary algorithm~\cite{bui2019greedy} for the same problem. This enhancement simplifies the current approach, facilitating the easier adoption of new policy language features. However, the search space of candidate policies explored by the existing evolutionary algorithm increases with the extension of policy language. That makes the algorithm's overall computation inefficient as it gives either the worst results or longer running time. The authors overcome this problem by adding a \emph{feature selection} phase to identify useful or essential features that helps to reduce the evolutionary algorithm's search space. 
Due to the flexibility and scalability of high-dimensional data and large datasets, the authors choose a neural network over other classification methods for the feature selection algorithm.
The algorithm maps different \emph{feature vectors} with Boolean values to indicate the authorization for users to act an action on resources.
Each \textit{feature} in the vector is defined as `a subject atomic condition, resource atomic condition, or atomic constraint satisfying the user-specified limits on lengths of paths in conditions and constraints'.

The authors train a 2-layer neural network that learns the classification of \emph{feature vectors}. The input layer contains various nodes where a node represents a feature in the feature vector. In contrast, the output layer has two different output nodes that determine the probabilities of classifying the input as deny or permit.
A feature vector is considered as \emph{unclassified} if there is an equal permit and deny probabilities. 
The training is continued for a fixed time limit, or terminates early if all inputs are correctly classified and no inputs are left unclassified.
The weights learned by the network are then analyzed to quantify \emph{how useful or necessary each feature is in determining the neural network's output}. Finally, the features are ordered using the score of each feature's contribution (usefulness) towards an output, and top-placed features are classified as the \emph{useful features}.
To demonstrate the effectiveness of the proposed method, the authors used the two ReBAC case studies~\cite{bui2019greedy} based on policies of real organizations~\cite{decat2014workforce, decat2014document} and some synthetic policies. 
Based on the evaluation, the algorithms that include \textit{feature selection} are significantly faster and achieve substantially better results than the algorithm without a feature selection stage.

The authors in~\cite{bui2020decision} proposed DTRM (Decision Tree ReBAC Miner) and DTRM$^-$, with the ability to mine policies in any ReBAC language.
The decision tree-based DTRM algorithm mines policies in the recent version of Object-oriented Relationship-based Access-control Language (ORAL) that supports two set comparison operators on top of existing operators in the earlier version. The DTRM$^-$ algorithm is an extension of DTRM with the support of negative conditions and negative constraints.
In general, the DTRM algorithm works in two main phases. First, it learns an authorization policy in a decision tree and then extracts a set of candidate authorization rules from the tree. 
Similar to~\cite{bui2019efficient}, the decision tree is trained to learn the classification of feature vectors, which are also mapped with Boolean values indicating the authorization for users to act on resources. The definition of a feature is the same as defined in~\cite{bui2019efficient}.
After building the tree, the DTRM algorithm converts the trained decision tree into an equivalent set of rules and includes them in the candidate policy.
In the second phase, first, the algorithm determines the mined policy by optionally eliminating negative conditions and constraints from the candidate rules.
This process is only applicable for the DTRM$^-$ algorithm. Next, the algorithm merges and simplifies the candidate rules.
The authors evaluate their proposed method using datasets from real organizations~\cite{decat2014workforce, decat2014document} and some synthetic policies~\cite{bui2019greedy}. 
Evaluation results show that the proposed method is effective at discovering the desired rules, producing smaller policies, and significantly faster than state-of-the-art approaches~\cite{iyer2019generalized, bui2019efficient}.

\begin{figure}[t]
\vspace{-2ex}
	\includegraphics[width=0.6\linewidth]{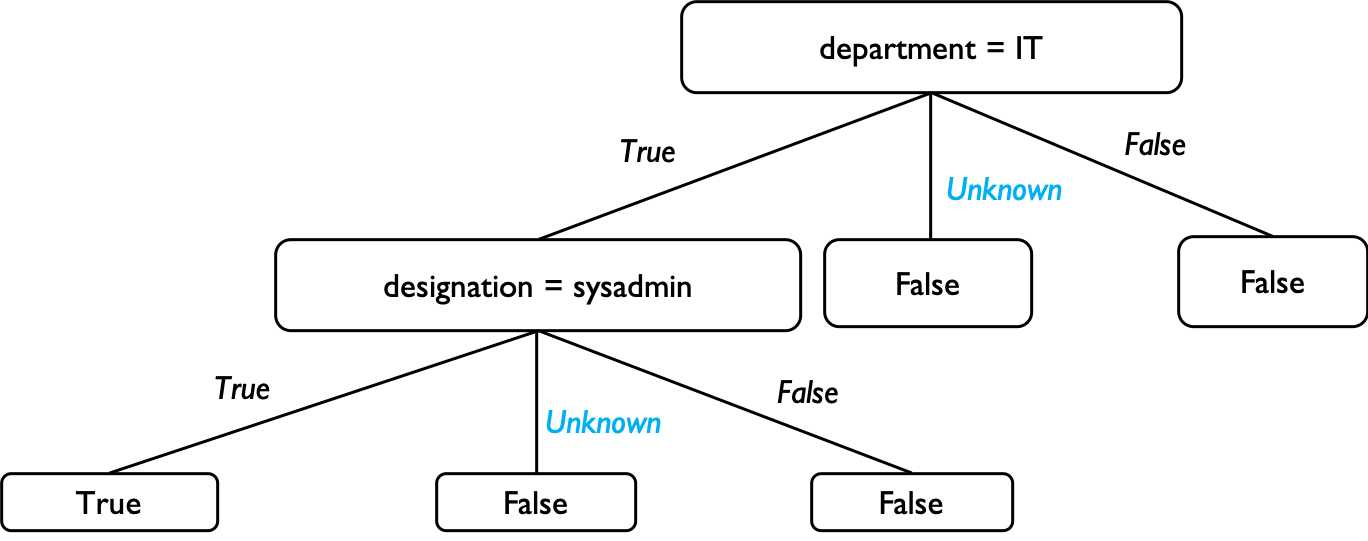}
	\caption{A multi-way decision tree with three truth values.}
	\label{fig:multiway_decision_tree}
	\vspace{-2ex}
\end{figure}

In most real-world data, information about permissions can be incomplete, or some attribute values can be missing (or unknown).
Authors in~\cite{bui2018mining, cotrini2018mining, law2020fastlas, iyer2020active} solved different variants of the ABAC and ReBAC policy mining problem considering \textit{incomplete permissions information}. However, all these works assume the attribute (and relationship in the case of ReBAC) information is complete (or known). The authors in~\cite{bui2020learning} proposed decision tree-based algorithms, Decision-Tree ReBAC Miner with Unknown values and negation (DTRMU$^-$ and DTRMU), for mining ABAC and ReBAC policies from ACLs. The first of its kind, the algorithm can deal with `incomplete' information about entities, where the values of some attributes of some entities are unknown.
To handle the `unknown' values, the authors proposed to reduce the core part of the ReBAC policy mining problem to the general problem of Kleene's three-valued logic learning formula. The authors assigned another truth value, `U,' to conditions and constraints with `unknown' values along with true (T) and false (F) binary logic values. The author also discussed that avoiding a three-valued logic by regarding them as false instead is unsafe.

To deal with a concise three-valued logic formula from a set of labeled feature vectors involving unknowns, the authors develop an algorithm based on multi-way decision trees (shown in Figure~\ref{fig:multiway_decision_tree}) as opposed to a binary decision tree. Identical to~\cite{bui2019efficient, bui2020decision}, the decision tree learns based on a given set of \emph{feature vectors} labeled with a permit or deny. The definition of feature is also same as~\cite{bui2019efficient} and~\cite{bui2020decision}.
The sets of features and feature vectors are extracted from an input ACL policy.
The learned decision tree is applied to classify authorization requests as permitted or denied and eventually create the tree's ReBAC rules (policy). 
The authors present their algorithm in the ReBAC mining context. However, to apply in the ABAC domain, one must limit the length of considered path expressions.
The authors experiment with their algorithm using four sample policies developed by Bui et al.~\cite{bui2019greedy} and two extensive case studies developed by Decat et al.~\cite{decat2014workforce, decat2014document} based on real-world applications. They didn't report any separate performance for the decision tree. Instead, the authors compare the algorithm's overall performance by comparing mined rules with simplified original policy rules.

\begin{table*}[t]
\centering
  \caption{Summarizing Machine Learning Based Policy Mining. The dataset type `RW', `RWA', and `Syn' indicates Real-World, Real-World Augmented, and Synthetic dataset, respectively. We link Dataset Type column to `SL.' column of Table~\ref{tab:real-world-dataset} if the respective dataset is public. We follow the same convention in Table~\ref{tab:verification},~\ref{tab:admin-monitoring}, and ~\ref{tab:access-decision}.}
  \label{tab:ACPolicyMining}
  \resizebox{\textwidth}{!}{%
  \begin{tabular}{cccccc}
    \hline
    \rowcolor{Gray}
    Reference
    & \multicolumn{1}{p{1.5cm}}{\centering Application}
    & \multicolumn{1}{p{4cm}}{\centering Problem Considered}
    & \multicolumn{1}{p{2cm}}{\centering Access\\Control\\Model}
    & \multicolumn{1}{p{3cm}}{\centering ML Approach}
    & \multicolumn{1}{p{3cm}}{\centering Dataset Type}\\\hline
    
    Frank et al. 2008~\cite{frank2008class} & 
    \multicolumn{1}{p{1.5cm}}{\centering Not specified} & 
    \multicolumn{1}{p{4cm}}{\centering Probabilistic bottom-up approaches for RBAC role mining} &
    \multicolumn{1}{p{2cm}}{\centering RBAC} &
    \multicolumn{1}{p{3cm}}{\centering Gibbs sampler \& Disjoint Decomposition} &
    \multicolumn{1}{p{3cm}}{\centering Syn \& RW}\\
    
    \rowcolor{LightGray}
    Ni et al. 2009~\cite{ni2009automating} & 
    \multicolumn{1}{p{1.5cm}}{\centering Not specified} & 
    \multicolumn{1}{p{4cm}}{\centering Adjustment of the mapping between roles and new privileges} & 
    \multicolumn{1}{p{2cm}}{\centering RBAC} & 
    \multicolumn{1}{p{3cm}}{\centering SVM (and others including DT, RF)} & 
    \multicolumn{1}{p{3cm}}{\centering Syn \& RW}\\
    
    Mocanu et al. 2015~\cite{mocanu2015towards} & 
    \multicolumn{1}{p{1.5cm}}{\centering Healthcare} & 
    \multicolumn{1}{p{4cm}}{\centering Policy inference from logs} & 
    \multicolumn{1}{p{2cm}}{\centering ABAC} & 
    \multicolumn{1}{p{3cm}}{\centering Restricted Boltzmann Machines} & 
    \multicolumn{1}{p{3cm}}{\centering Syn}\\
    
    \rowcolor{LightGray}
    Benkaouz et al. 2016~\cite{benkaouz2016work} & 
    \multicolumn{1}{p{1.5cm}}{\centering Not specified} & 
    \multicolumn{1}{p{4cm}}{\centering Classification and clustering of policies} & 
    \multicolumn{1}{p{2cm}}{\centering ABAC} &  
    \multicolumn{1}{p{3cm}}{\centering K-Nearest Neighbors}  & 
    \multicolumn{1}{p{3cm}}{\centering Not Used}\\
    
    Narouei et al. 2017~\cite{narouei2017towards} & \multicolumn{1}{p{1.5cm}}{\centering Not specified} & 
    \multicolumn{1}{p{4cm}}{\centering Policy extraction from natural language documents}  & 
    \multicolumn{1}{p{2cm}}{\centering ABAC} & 
    \multicolumn{1}{p{3cm}}{\centering Recurrent Neural Network} & 
    \multicolumn{1}{p{3cm}}{\centering Syn}\\
    
    \rowcolor{LightGray}
    El Hadj et al. 2017~\cite{el2017abac} & 
    \multicolumn{1}{p{1.5cm}}{\centering Not specified} & 
    \multicolumn{1}{p{4cm}}{\centering Classification and clustering of policies} & 
    \multicolumn{1}{p{2cm}}{\centering ABAC} &  
    \multicolumn{1}{p{3cm}}{\centering K-Nearest Neighbors}  & 
    \multicolumn{1}{p{3cm}}{\centering Syn}\\
    
    Alohaly et al. 2018~\cite{alohaly2018deep}  & 
    \multicolumn{1}{p{1.5cm}}{\centering Not specified} & 
    \multicolumn{1}{p{4cm}}{\centering ABAC attribute extraction from natural language} &
    \multicolumn{1}{p{2cm}}{\centering ABAC} &
    \multicolumn{1}{p{3cm}}{\centering CNN} &
    \multicolumn{1}{p{3cm}}{\centering RWA (SL: \ref{row:ibm}, \ref{row:iTrust}, \ref{row:cyber-chair}, \ref{row:coll-acp})}\\\\
    
    \rowcolor{LightGray}
    Karimi et al. 2018~\cite{karimi2018unsupervised} & \multicolumn{1}{p{1.5cm}}{\centering Healthcare, education} & 
    \multicolumn{1}{p{4cm}}{\centering Policy extraction} & 
    \multicolumn{1}{p{2cm}}{\centering ABAC} & 
    \multicolumn{1}{p{3cm}}{\centering K-modes} & 
    \multicolumn{1}{p{3cm}}{\centering Syn}\\\\
    
    Cotrini et al. 2018~\cite{cotrini2018mining} & 
    \multicolumn{1}{p{1.5cm}}{\centering Not specified} & 
    \multicolumn{1}{p{4cm}}{\centering Policy mining} & 
    \multicolumn{1}{p{2cm}}{\centering ABAC} & 
    \multicolumn{1}{p{3cm}}{\centering APRIORI-SD~\cite{kavvsek2006apriori}} & 
    \multicolumn{1}{p{3cm}}{\centering Syn \& RW (SL: \ref{row:amazon-uci}, \ref{row:amazon-kaggle})}\\\\
    
    \rowcolor{LightGray}
    Alohaly et al. 2019~\cite{alohaly2019automated}  & 
    \multicolumn{1}{p{1.5cm}}{\centering Not specified} & 
    \multicolumn{1}{p{4cm}}{\centering Attribute extraction from natural language for hierarchical ABAC} &
    \multicolumn{1}{p{2cm}}{\centering ABAC} &
    \multicolumn{1}{p{3cm}}{\centering CNN} &
    \multicolumn{1}{p{3cm}}{\centering RWA (SL: \ref{row:ibm}, \ref{row:iTrust}, \ref{row:cyber-chair}, \ref{row:coll-acp})}\\
    
    Alohaly et al. 2019~\cite{alohaly2019towards}  & 
    \multicolumn{1}{p{1.5cm}}{\centering Not specified} & 
    \multicolumn{1}{p{4cm}}{\centering ABAC constraints extraction from natural language policies} &
    \multicolumn{1}{p{2cm}}{\centering ABAC} &
    \multicolumn{1}{p{3cm}}{\centering BiLSTM} &
    \multicolumn{1}{p{3cm}}{\centering RWA}\\\\
    
    \rowcolor{LightGray}
    Zhou et al. 2019~\cite{zhou2019automatic}   & 
    \multicolumn{1}{p{1.5cm}}{\centering SCADA} & 
    \multicolumn{1}{p{4cm}}{\centering Role and permission assignments} &
    \multicolumn{1}{p{2cm}}{\centering RBAC} &
    \multicolumn{1}{p{3cm}}{\centering SVM \& Adaboost} &
    \multicolumn{1}{p{3cm}}{\centering RWA (SL: \ref{row:scada-intrusion})}\\\\
    
    Bui et al. 2019~\cite{bui2019efficient} & 
    \multicolumn{1}{p{1.5cm}}{\centering Not specified} & 
    \multicolumn{1}{p{4cm}}{\centering Policy mining} & 
    \multicolumn{1}{p{2cm}}{\centering ReBAC} & 
    \multicolumn{1}{p{3cm}}{\centering Neural Network} & 
    \multicolumn{1}{p{3cm}}{\centering Syn \& RW (SL: \ref{row:e-document}, \ref{row:work-force})}\\\\
    
    \rowcolor{LightGray}
    Bui et al. 2020~\cite{bui2020decision} & 
    \multicolumn{1}{p{1.5cm}}{\centering Not specified} & 
    \multicolumn{1}{p{4cm}}{\centering Policy mining} & 
    \multicolumn{1}{p{2cm}}{\centering ReBAC} & 
    \multicolumn{1}{p{3cm}}{\centering Decision Tree} & 
    \multicolumn{1}{p{3cm}}{\centering Syn \& RW (SL: \ref{row:e-document}, \ref{row:work-force})}\\\\
    
    Bui et al. 2020~\cite{bui2020learning} & 
    \multicolumn{1}{p{1.5cm}}{\centering Not specified} & 
    \multicolumn{1}{p{4cm}}{\centering Policy mining} & 
    \multicolumn{1}{p{2cm}}{\centering ABAC, ReBAC} & 
    \multicolumn{1}{p{3cm}}{\centering Decision Tree} & 
    \multicolumn{1}{p{3cm}}{\centering Syn \& RW (SL: \ref{row:e-document}, \ref{row:work-force})}\\\\

    \rowcolor{LightGray}
    Jabal et al. 2020~\cite{jabal2020polisma} & 
    \multicolumn{1}{p{1.5cm}}{\centering Not specified} & 
    \multicolumn{1}{p{4cm}}{\centering Learn ABAC policies from logs of historical access requests and their corresponding decisions} & 
    \multicolumn{1}{p{2cm}}{\centering ABAC} & 
    \multicolumn{1}{p{3cm}}{\centering RF, KNN} & 
    \multicolumn{1}{p{3cm}}{\centering Syn \& RW (SL: \ref{row:amazon-uci})}\\\\

    Karimi et al. 2021~\cite{karimi2021automatic} & \multicolumn{1}{p{1.5cm}}{\centering Not specified} & 
    \multicolumn{1}{p{4cm}}{\centering Automating ABAC policy extraction based on access logs} & 
    \multicolumn{1}{p{2cm}}{\centering ABAC} & 
    \multicolumn{1}{p{3cm}}{\centering K-modes} & 
    \multicolumn{1}{p{3cm}}{\centering Syn \& RW (SL: \ref{row:amazon-uci}, \ref{row:amazon-kaggle})}\\\\
    
    \rowcolor{LightGray}
    Heaps et al. 2021~\cite{heaps2021access} & 
    \multicolumn{1}{p{1.5cm}}{\centering Not specified} & 
    \multicolumn{1}{p{4cm}}{\centering Extracting access control policy from user stories} & 
    \multicolumn{1}{p{2cm}}{\centering RBAC and ABAC} & 
    \multicolumn{1}{p{3cm}}{\centering Transformers, CNN, SVM} & 
    \multicolumn{1}{p{3cm}}{\centering RW (SL: \ref{row:dalpiaz})}\\
    
    \hline
  \end{tabular}
  }
\end{table*}

\subsection{Policy Verification and Testing}

\subsubsection{Policy Verification and Testing}

\begin{figure}[t]
    \centering
	\fbox{\includegraphics[width= 0.5\linewidth, scale = 0.6]{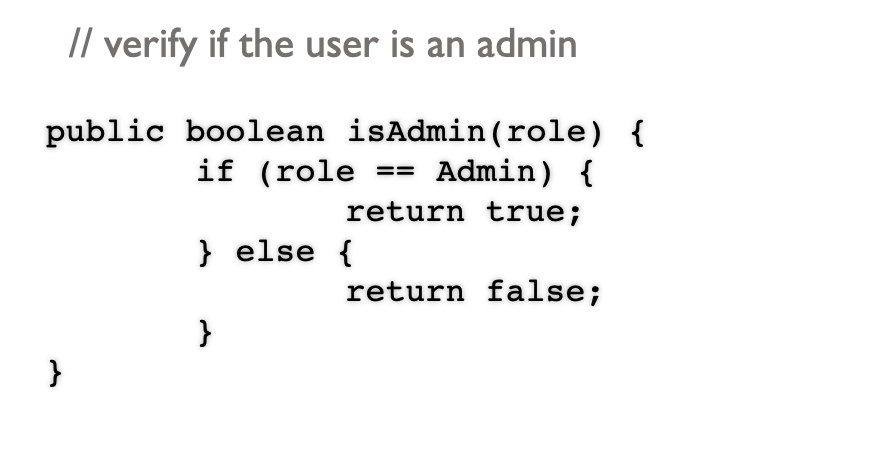}}
	\caption{The source code of a function that verifies a user as Admin.}
	\label{fig:codeSnippetFunction}
	\vspace{-2ex}
\end{figure}

Traditional policy verification methods are infeasible for many systems as they are error-prone and time-consuming processes. These methods perform static or dynamic analysis on the access control system to determine the policy behavior. Such analysis techniques can not specify roles (as shown in Figure~\ref{fig:codeSnippetFunction}) or permissions-specific code elements, find relations among these codes and associated policy elements, or tackle the case when a mapping is required for a new code and policy element. 
The authors in~\cite{heaps2019toward} discuss these issues and pretend a potentiality of developing a more robust and efficient system by leveraging recent advancements in deep learning. The authors propose to train a deep learning model based on the links among the code and policy elements. However, the model can not learn anything based on the input code elements since they do not have a numerical meaning. Therefore, embedding code elements and mapping the word into a high-dimensional space are required to acquire meaningful numerical information to expedite learning. The authors utilize the popular Skip-gram~\cite{mikolov2013efficient} word embedding algorithm Word2Vec and adjust the algorithm to obtain word embedding for code elements in the source code. The authors experiment proposed technique on the Java Development Kit 8 (JDK8)\footnote{www.oracle.com/technetwork/java/javase/overview/index.html} and Apache Shiro\footnote{https://shiro.apache.org/} access control library. The evaluation results signify that the intended technique can produce high-quality word embedding and deliver state-of-the-art performance.

Access control policies are verified based on model proof, data structure, system simulation, and test oracle to ensure the policy works desirably. By default, it is assumed that the policy under verification is accurate unless the policy failed to hold for test cases. However, this comprehensive test case generation is challenging and somewhat impractical. To make the entire verification process more efficient and straightforward, in a NIST Internal Report (IR), Vincent C. Hu~\cite{hu2021verificationNIST} propose a technique exploiting machine learning algorithm. Compared to traditional methods, the proposed method, which is feasible and efficient, can check the policy logic directly. For the verification task, a machine learning model is trained based on the values of the assigned attributes of policy rules to generate a classification model. The trained model will be used to predict the access permissions (e.g., grant, deny, etc.) assigned to various rules that allow for the detection of inconsistencies, indicating faults found among policy rules. The author applies RF as the ML method in the verification process. As part of training data preparation, it is required to encode access control policy rules in a data table. Each column contains an attribute, action, or permission value, and the row represents a policy rule. The access permission is determined based on the attribute and action values in the row. The data table contains enforceable rules only that ensure the \textit{syntactical correctness} of a policy.

Besides, any rule with more than one action or object attribute value is split into separate sub-rules to hold only one action and one object attribute value to comply with the RF tree node representing an attribute or action. After training an RF model, the evaluation is done to detect permission conflict rules and ensure that the model can recognize different policy rules semantic that include condition property, separation of duty property, and exclusion property. The analysis of the accuracy function provides the percentage of \textit{semantic correctness} of the original policy versus the RF model. Correctness of less than 100\% indicates that conflict (error) rules may exist in the policy. Overall, the algorithm can efficiently verify the policy and detect inconsistencies (i.e., faults) in the policy rules.
Table~\ref{tab:verification} summarizes
access control policy verification tools and methods using machine learning.


\begin{table*}[t]
\centering
  \caption{Summarizing Machine Learning Based Policy Verification. `Common': any access control model.}
  \resizebox{\textwidth}{!}{%
  \begin{tabular}{cccccc}
    \hline
    \rowcolor{Gray}
    Reference
    & Application
    & \multicolumn{1}{p{5.5cm}}{\centering Problem Considered}
    & \multicolumn{1}{p{1.5cm}}{\centering Access Control Model}
    & \multicolumn{1}{p{2.5cm}}{\centering ML Approach}
    & \multicolumn{1}{p{2.5cm}}{\centering Dataset Type}\\
    \hline

    Heaps et al. 2019~\cite{heaps2019toward} & Not Specified & \multicolumn{1}{p{5.5cm}}{\centering Word embedding technique for developing automated policy verification tools} & RBAC & Neural Network & Syn\\
    
    \rowcolor{LightGray}
    Vincent C. Hu 2021~\cite{hu2021verificationNIST} & Not Specified & \multicolumn{1}{p{5.5cm}}{\centering Verifying access control policy} & Common & RF & Syn\\
    
    \hline
  \end{tabular}
  }
  \label{tab:verification}
\end{table*}

\subsection{Policy Administration and Monitoring}
Table~\ref{tab:admin-monitoring} outlines related methods proposed for the access control administration and monitoring using machine learning.

\subsubsection{Policy Administration}

Manual access policy updates are a laborious and error-prone task~\cite{argento2018towards} and the system becomes vulnerable to different cyber threats.
Authors in~\cite{argento2018towards} present an ML-based approach, named ML-AC, that can update policies at run-time automatically and prevent such threats. The ML-AC monitor and learn the \textit{access behavioral features} (e.g., frequency of access, amount of data, location, etc.) of a user and adjust existing access control rules based on learned \textit{contextual knowledge}. The authors apply this method by adding a novel \textit{Contextual Behaviour Learning} component in the Policy Administration Point (PAP) of the access control system. This new component builds user profiles based on the monitored access pattern and adjusts access control policies by utilizing those profiles.
The authors introduce the idea of \textit{user behavior} that indicates how users are utilizing resources and build a behavioral model to identify anomalous accesses. The behaviors are formally defined in access request attributes (e.g., user, resource, and action) and different contextual features (i.e., working time, location, types of activities, etc.). Next, ML-AC clusters such behaviors together based on uniformity and create various behavioral profiles of normal accesses. 

The ML-AC uses RF to classify a given access control behavior as the \textit{normal} or \textit{anomalous}. To refine the existing policy, ML-AC obtains the output of the decision tree, defined as \textit{ML-rules}. These ML rules are encoded in existing policy in additional conditions that transfer learned contextual knowledge of user behavior. Such processed rules have more control in determining abnormal accesses in the system. To keep the RF models accurate and updated, the ML-AC also monitors the evolution of user behaviors to detect the appearance of new clusters using a method Olindda~\cite{spinosa2009novelty}.
The authors implement ML-AC based on a synthetic dataset having 3000 unique behaviors from normal and anomalous categories to evaluate the effectiveness and practicality. The method presents the behaviors as numerical features based on three-dimensional points. The source code and dataset are publicly available in GitHub\footnote{https://github.com/cybersoton/ml-ac}. The authors compare ML-AC with a method name BBNAC~\cite{frias2009network} which is similar to ML-AC. The authors also compare the performance with a variant of ML-AC named ML-AC$_{nok}$ where they did not use any contextual knowledge. The experimentation shows ML-AC performs significantly better in all the cases.


The authors in~\cite{alkhresheh2020adaptive} propose an adaptive access control policy framework for IoT deployments. The framework dynamically refines the access policies based on the access behaviors of the IoT devices. In parallel to the traditional ABAC authorization server, the authors propose incorporating a policy management module that implements the access policy adaptation functionalities, including the access behavior classifier and the policy refinement components. For both servers, a context monitor component is added to access the contextual information.
The authors apply RF, and Recurrent Neural Networks (RNNs)~\cite{zaremba2014recurrent} on three years of access history data of a door locking system of a university to justify the effectiveness of the proposed method. The authors did not publish their dataset. For RNN, the authors use Long Short-Term Memory (LSTM)~\cite{hochreiter1997long} due to its efficiency for large data sets. The authors create different instances of the dataset based on the size. The empirical results imply that both RF and LSTM show similar performance for smaller-sized datasets. For larger instances, the LSTM outperforms RF because of the increased chances for the LSTM classifier to learn from longer access sequences. Therefore, the authors conclude that the LSTM classifier can scale better to IoT environments.


Gumma et al.~\cite{gumma2021pammela} propose PAMMELA which is an ML-based ABAC policy administration method. The PAMMELA creates new rules for the proposed changes and extends existing policy by adjusting the newly developed rules with the current policy. PAMMELA can also develop new policies for a system by learning existing policy rules in a similar system.
PAMMELA is a supervised ML-based approach that mainly works in two phases. In the first phase, PAMMELA trains an ML classifier on ABAC policy rules. The authors define a rule as the combination of subject and object attribute-value pairs indicating whether the subject has specific access or not to the associated object. In the second phase, PAMMELA provides a set of access requests to the trained classifier and creates a set of rules based on the classifier's access decision for respective requests. These generated rules capture the access control state of the given ABAC policy rules. The authors experiment with their proposed method using three manually created datasets with various ABAC rules, subject attributes values, object attribute values, and access requests. To evaluate the performance of PAMMELA, the authors apply different kinds of machine learning methods, including neural network, decision tree, RF, Extra Trees (ET)~\cite{geurts2006extremely}, Gradient Boosting (GB)~\cite{friedman2001greedy} and Extreme Gradient Boosting (XGB)~\cite{chen2015xgboost}. For a proper justification of the proposed method, the authors follow multiple evaluation techniques. The authors highlight some insights on managing ABAC policies using PAMMELA based on empirical results.

\subsubsection{Policy Monitoring}
Due to the lack of proper tools, standardized policy specifications such as XACML and their implementation can be error-prone. It demands rigorous verification and validation to ensure the implemented policy complies with the desired one. Martin et al.~\cite{martin2006inferring} discover that the ML algorithms could summarize the basic properties of such a policy and help to identify specific bug-exposing requests. Especially, their proposed method can efficiently identify the discrepancies between the policy specification and its intended functionalities.
The authors first generate a set of access requests and apply them to the existing access control system to observe the behavior of the policy. These observations are structured as request-response pairs and given as input to the ML algorithm to summarize the policy in the form of inferred properties. Essentially, these inferred properties capture the general policy behavior to determine cases that might be bug-exposing and necessitate a manual inspection. The authors integrate Sun’s XACML implementation~\cite{SunXacml2005} and Weka~\cite{ian2005data}, a collection of ML algorithms for data mining tasks, into their proposed tool that implements the entire approach through request generation, request evaluation, and policy property inference. Remarkably, the authors use the Prism classification algorithm~\cite{cendrowska1987prism} for the rule sets generation. For policy verification, the tool is applied to an access control policy of a central grades repository system for a university~\cite{fisler2005verification}. Eventually, the generated rules are translated into an XACML policy to evaluate the performance. Evaluation results show that the inferred properties can indeed summarize the given policy that supports the ability of ML in providing crucial insight into basic policy properties, and help identify specific bug-exposing requests.

\begin{figure}[t]
    \vspace{-2ex}
	\includegraphics[width=0.6\linewidth]{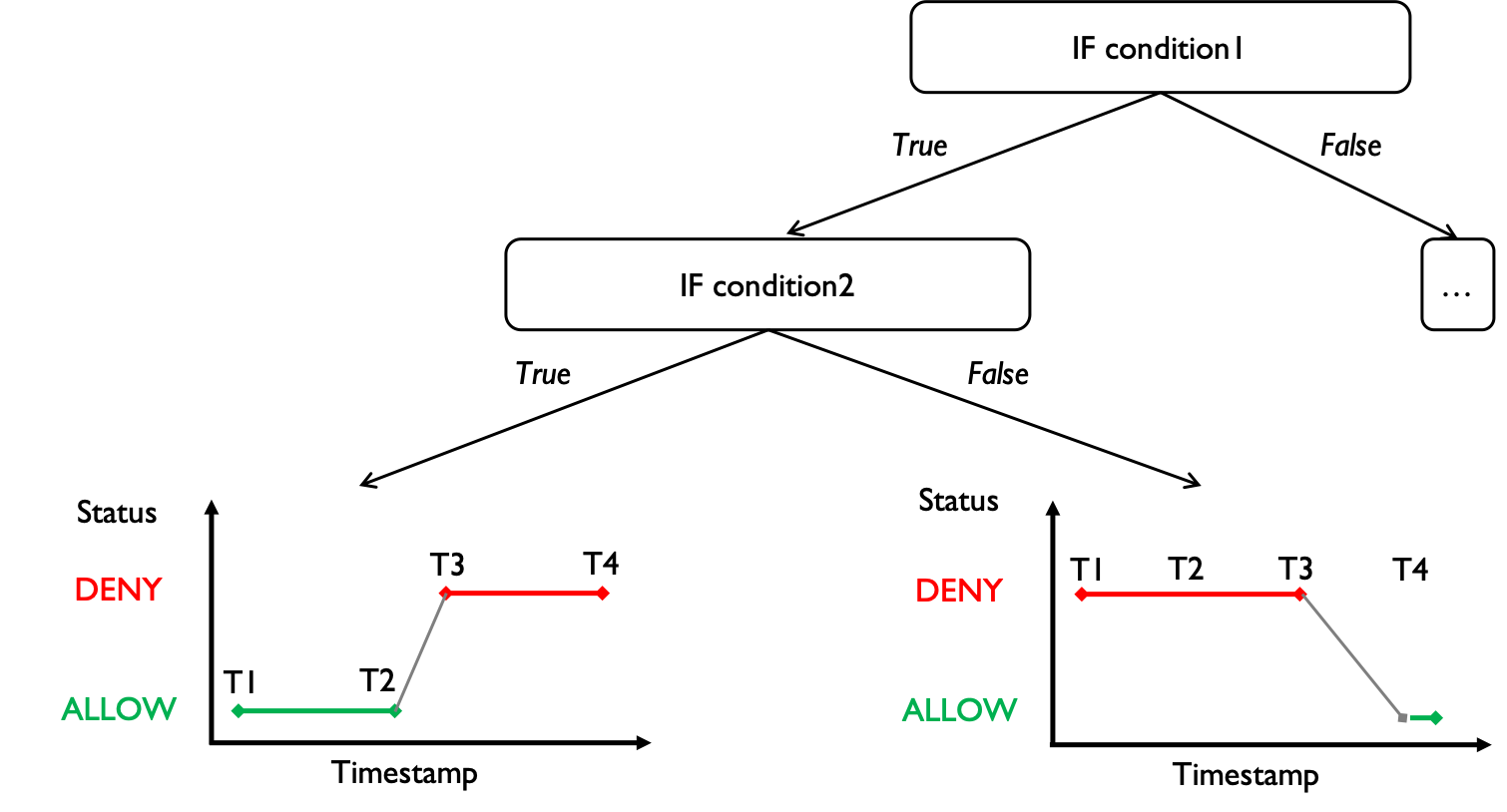}
	\caption{Time-Changing Decision Tree (TCDT).}
	\label{fig:time_changing_decision_tree}
	\vspace{-3mm}
\end{figure}

A regular access control policy update is required to accommodate intermediate access changes. The task is very challenging for system admins and error-prone (e.g., over-granting access privilege) due to the lack of proper tools~\cite{xiang2019towards}. However, failing to detect such requests (or misconfigurations) in the early stage may cause severe security incidents.
Authors in~\cite{xiang2019towards} develop a tool named P-DIFF to backing system admins to monitor access control policy changes. P-DIFF also supports investigating any malicious access by backtracking related changes that help system admins identify the reason behind unwanted access. The authors adopt a decision tree representation to convert heterogeneous access control behaviors to a standard rule-based format. However, a regular decision tree is limited in handling time-series information to consider access concerning time. To compensate for such a challenge, the authors extend the decision tree algorithm and propose a novel Time-Changing Decision Tree (TCDT). Unlike single binary value outcome (allow or deny), each rule of a TCDT is expressed as a time series as depicted in Figure~\ref{fig:time_changing_decision_tree}. The TCDT learning algorithm is designed to infer the tree by considering access logs as a sequence of access events ordered by access time. Then, the TCDT allows modeling access control behavior at any given time along with monitoring changes in the access control rules. 

The P-DIFF takes access logs as the input to generate a TCDT as the output that records the entire change history to aid both access policy change validation and forensics analysis.
The authors experiment effectiveness of P-DIFF across datasets from five real-world systems, including the access logs from a public dataset of request traces to Wikipedia~\cite{urdaneta2009wikipedia}. Experiments demonstrate that P-DIFF can detect 86\%$-$100\% of access control policy changes with an average precision of 89\%. In forensic analysis, P-DIFF shows 85\%$-$98\% efficacy in determining the root cause of several cases.

\begin{table*}[t]
\centering
  \caption{Summarizing Machine Learning Based Policy Administration and Monitoring. `Common': any access control model.}
  \resizebox{\textwidth}{!}{%
  \begin{tabular}{cccccc}
    \hline
    \rowcolor{Gray}
    Reference
    & Application
    & \multicolumn{1}{p{4cm}}{\centering Problem Considered}
    & \multicolumn{1}{p{2cm}}{\centering Access Control Model}
    & \multicolumn{1}{p{2.2cm}}{\centering ML Approach}
    & \multicolumn{1}{p{2.2cm}}{\centering Dataset Type}\\\hline
    
    Martin et al. 2006~\cite{martin2006inferring} & Not Specified & \multicolumn{1}{p{4cm}}{\centering Inferring policy and identifying bug-exposing requests} & 
    \multicolumn{1}{p{2cm}}{\centering Common} & 
    \multicolumn{1}{p{2.2cm}}{\centering Prism~\cite{cendrowska1987prism}} & 
    \multicolumn{1}{p{2.2cm}}{\centering RW (SL:~\ref{row:university-data})}\\
    
    \rowcolor{LightGray}
    Argento et al. 2018~\cite{argento2018towards}  & 
    Not Specified & 
    \multicolumn{1}{p{4cm}}{\centering Improves the policy administration point (PAP) of the ABAC model} &
    \multicolumn{1}{p{2cm}}{\centering ABAC (and Common)}  &
    \multicolumn{1}{p{2.2cm}}{\centering RF and K-means Clustering} &
    \multicolumn{1}{p{2.2cm}}{\centering Syn}\\
    
    Xiang et al. 2019~\cite{xiang2019towards} & Not Specified & \multicolumn{1}{p{4cm}}{\centering Continuous access control validation and forensics} & 
    \multicolumn{1}{p{2cm}}{\centering Common} & 
    \multicolumn{1}{p{2.2cm}}{\centering Time-Changing Decision Tree (TCDT)} & 
    \multicolumn{1}{p{2.2cm}}{\centering RW (SL:~\ref{row:wikipedia})}\\
    
    \rowcolor{LightGray}
    Ashraf et al. 2020~\cite{alkhresheh2020adaptive}  & 
    IoT & 
    \multicolumn{1}{p{4cm}}{\centering Refines the access policies in response to changes in the access behavior} &
    ABAC  &
    \multicolumn{1}{p{2.2cm}}{\centering RF and RNN} &
    \multicolumn{1}{p{2.2cm}}{\centering RW}\\
    
    Gumma et al. 2021~\cite{gumma2021pammela}  & 
    Not Specified & 
    \multicolumn{1}{p{4cm}}{\centering ABAC policy administration} &
    ABAC &
    \multicolumn{1}{p{2.2cm}}{\centering Neural Network, DT, RF, etc.} &
    \multicolumn{1}{p{2.2cm}}{\centering Syn}\\
    
    \hline
  \end{tabular}
  }
  \label{tab:admin-monitoring}
\end{table*}

\subsection{ML for Access Control}

\subsubsection{Access Decision}
Contemporary researches manifest the advantages of using an ML model for more accurate access control decision-making~\cite{karimi2021adaptive,cappelletti2019quality,chang2006access,liu2021efficient,nobi2022toward,srivastava2020machine}. These systems decide accesses based on a trained ML model instead of a written access control policy.
Generally, these models make access control decisions (grant or deny) using user and resource metadata and attributes. Metadata and attributes are the user/resource features that an ML model learns for subsequent access decisions. We briefly discuss these approaches below and summarize them in Table~\ref{tab:access-decision}.

There are situations in various systems where access control policies need to be restricted with access hours and established policies.
For example, a user has access to a resource, but there is a policy that the user can not access the respective resource outside of office hours.
Chang et al.~\cite{chang2006access} proposed a novel access control system, called time-constraint access control, that can be applied to access policies constrained with time.
The authors apply SVM to the proposed scheme and divide the processes into three phases: (1) the input pattern transforming, (2) the training phase for SVMs, and (3) the authority decision phase. The authors implemented SVMs using the Library of Support Vector Machines (LIBSVM)~\cite{chang2011libsvm}, which is an efficient and easy-to-use software implementing support vector learning. As part of training data, the system administrator determines a login time and a password for each user. The trained SVMs can classify the users into their groups and give them corresponding security access using their passwords and system login time. Consequently, instead of written access control policies, trained SVMs are used for access decisions. The authors evaluate the performance of access decisions with the same data used for the training. The experimental results show that the system can authenticate users' access rights, which justifies the practicality of the proposed approach.

In general, centralized access control architectures with static security policies are limited in the IoT domain. The IoT devices don't have intensive and computational capabilities or storage to deal with a complete distributed access control process. However, outsourcing access control management to non-constrained nodes is not practical as it might introduce severe security and privacy concerns. As a result, IoT needs an access control framework suitable to its distributed nature, where users may control their privacy, and at the same time, the need arise for centralized entity handling. The author in~\cite{outchakoucht2017dynamic} proposed a blockchain and machine learning-based access control model for the IoT domain. As presented, the distributed model will have distributed policy handled with blockchain, which is consistent with the decentralized aspect of IoT. Also, the authors proposed to utilize the power of AI, especially those of machine learning, to facilitate the adjustments in the access control policy dynamically. In particular, they proposed applying Reinforcement Learning due to its ability to adjust the knowledge incrementally while accesses are made to resources, and the security policy is executed. The authors didn't present any implementation of their work or case study as they kept it for future work. 

Luca Cappelletti et al~\cite{cappelletti2019quality} investigate different ML techniques from both symbolic (Decision Tree (DT), Random Forest (RF)) and non-symbolic (Support Vector Classifier (SVC) and Multi-Layer Perceptron (MLP)) families for inferring ABAC policies from access logs. The experimentation is performed on two Amazon Datasets~\cite{AmazonUCI2011, AmazonKaggle2013} and the Incident dataset~\cite{amaral2018enhancing}. The details of the datasets are reported in SL:~\ref{row:amazon-uci}, \ref{row:amazon-kaggle}, and \ref{row:incident} of Table~\ref{tab:real-world-dataset}.
The Amazon datasets consist of access logs that indicate the decisions (grant or deny) for numerous accesses requested by thousands of Amazon users. Hence, the data is sparse and imbalanced in the sense that it has extensive logs with \emph{grant} access. The Incident dataset, on the other hand, contains more balanced data from the different classes. 
The authors further justify these characteristics using PCA and t-SNE~\cite{xyntarakis2019data} data visualization of some instances of all these datasets. Both PCA and t-SNE demonstrate that the classes in the Incident log form well-separated clusters. In contrast, the negative class in Amazon dataset does not form an easily separable group. That makes it challenging to create straightforward and interpretable rules for Amazon datasets. 

As part of Amazon datasets preprocessing, the columns with either many distinct classes or a single class are removed. Also, the method applies one-hot encoding to the original data before applying them to the ML algorithms as input.
For the experimentation on Amazon datasets, a variant of the balanced Monte Carlo cross-validation method was adopted for training and evaluation. As far as performance is concerned, the accuracy for Amazon datasets of all the techniques is about 50\%, which is expected due to the inadequacy of typical ML techniques to handle imbalance and sparse data.
On the contrary, the accuracy is exceptionally high for the balanced and non-sparse Incident logs. However, in the case of sparse and imbalanced logs, among all the techniques, MLP performed better to demonstrate the ability of neural networks to capture complicated relationships in the data. Also, the policies obtained from symbolic ML techniques provide a better understanding of a decision than the non-symbolic peers. As a result, the authors raised the concern of \textit{incompatibility} of such \emph{black-box} ML techniques for the domains such as the ABAC system, where \emph{explainability} and \emph{verificability} are important.

\begin{figure}[t]
    \vspace{-2ex}
	\includegraphics[width=0.8\linewidth]{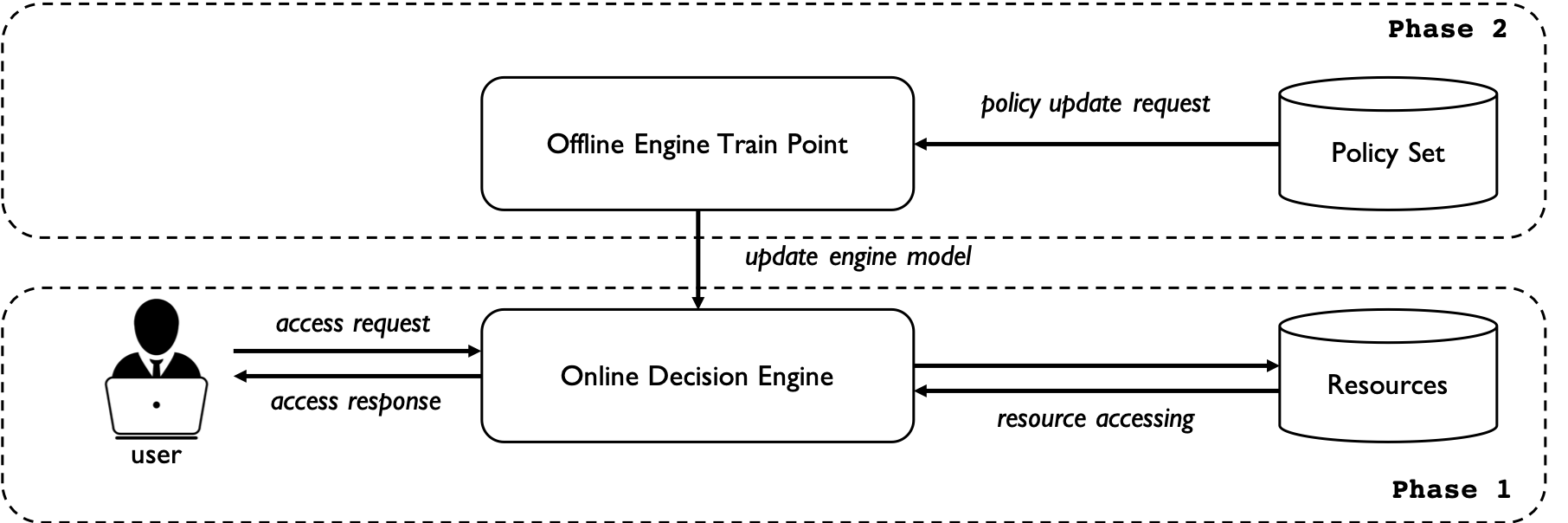}
	\caption{Permission Decision Method of EPDE-ML.}
	\label{fig:epde_ml_permission_decision}
	\vspace{-2ex}
\end{figure}

Over time, various access control models such as cryptographic-based, identity-based, and trust-based access control models have been proposed for secure access in the cloud environment. However, such models overlook the user behavior and scalability of the trust management system, which are essential factors for cloud-based systems. Khilar et al.~\cite{khilar2019trust} propose a trust-based approach that allows access to the cloud resources based on the accesses history and access behavior of the user. Also, the approach considers other factors, including user behavior, bogus request, unauthorized request, forbidden request, and range specifications.
To justify the efficacy of their work, the authors experiment with several ML techniques such as KNN, decision tree, logistic regression, naive Bayes, neural network, etc. They also consider different ensemble algorithms for the evaluation. They evaluate the performance of users and resources separately. The ensemble model of random forest and K-nearest neighbor performs better than other models. Overall, the neural network-based models achieve the highest performance, demonstrating the neural network's superior ability to capture access control relationships.

There are many real-world applications such as defense, airport surveillance, and hospital management systems. The legacy access control systems are ineffective as the policy decided initially cannot be changed dynamically. A situation may arise when a user needs to access some resources even though the current policy doesn't support the access. However, one could consider granting access considering the nature of applications, the authenticity, and the requesters' need. Srivastava et al.~\cite{srivastava2020machine} propose a novel access control framework named risk adaptive access control (RAdAC), which understands the genuineness of the requester, calculates the risk, and then acts accordingly. The framework considers many real-world attributes in its design, such as time of access, location of access, frequency of the requests, and the sensitivity of requested resources. The proposed approach depends on multiple parameters such as a sensitivity score and relevance, and the authors followed some unique strategies to engineer those parameters.
The authors develop a prototype of their proposed method for a Hospital Management System (HMS). They experiment with a neural network with two hidden layers and an RF algorithm. For RF, they examine two cases. In one case, they use the input data without engineered parameters (e.g., sensitivity score) and apply the autoencoder~\cite{farahnakian2018deep} to extract features from the data. In another case, they use both input data and the engineered parameters. The RF achieves the best results among all these approaches using both input data and parameters. However, based on evaluation, the authors confirm that domain expertise is required for determining the optimal values of different parameters.


The authors in ~\cite{liu2021efficient} propose an Efficient Permission Decision Engine scheme based on Machine Learning (EPDE-ML) that improves the policy decision point (PDP) of the ABAC model.
Internally, the EPDE-ML includes an RF algorithm trained based on user attributes and prior access control information. For any access control request from a user, the decision engine results either permit or deny indicating the user has access or not to the corresponding resource. 
The EPDE-ML splits the traditional permission decision process into two phases (Phase 1 and Phase 2), shown in Figure~\ref{fig:epde_ml_permission_decision}. 

In Phase 2, to ensure the privacy protection of the policy information, the RF model is trained based on the current access control policy information. The trained model is used to predict access control decisions in Phase 1. In the case of policy change, the method updates the RF model in Phase 2 and replaces the model in Phase 1 after that.
The authors perform experiments on Amazon's real access control policy set~\cite{AmazonKaggle2013} and compare the results with some other ML algorithms. The authors evaluate the performance of different approaches based on several metrics that include accuracy, precision, recall, and F1-score. The author also analyzes the receiver operating characteristic (ROC) curve and area under the curve (AUC) among all the methods. Overall, the EPDE-ML demonstrates superior performance with an AUC value of 0.975 and an accuracy of 92.6\%. Furthermore, the authors also compare the decision time that shows the EPDE-ML requires a consistent decision time of around 0.115 seconds regardless of the policy size.

\begin{figure}[t]
    \vspace{-3ex}
	\includegraphics[width=0.5\textwidth]{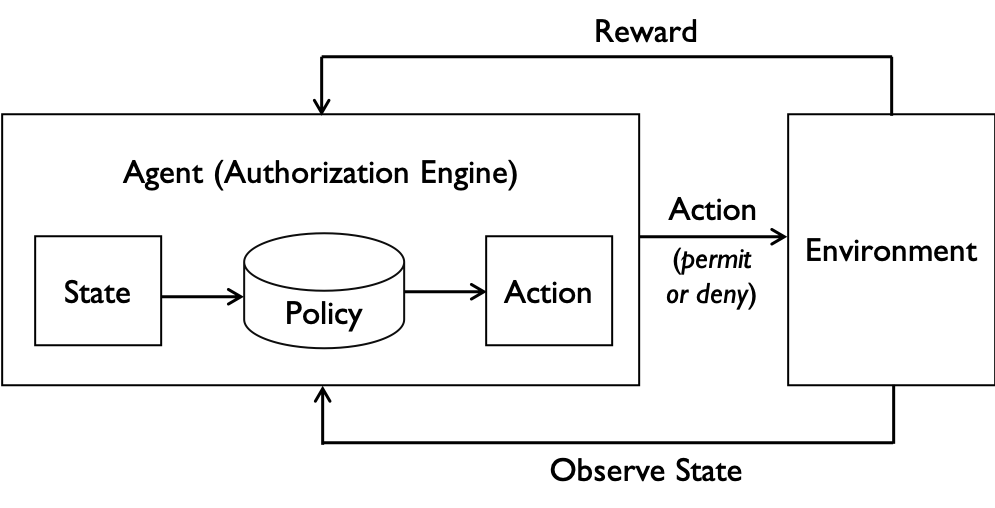}
	\caption{Overview of Reinforcement Learning Based ABAC Policy Learning.}
	\label{fig:adaptive_rl_policy_learning}
	\vspace{-3ex}
\end{figure}

There are some challenges with supervised learning approaches in the real-world system, such as limited labeled data, sparse logs holding partial access information, etc. To tackle these issues, Karimi et al.~\cite{karimi2021adaptive} propose an adaptive access control approach that learns from the feedback provided by the user.
The authors propose a reinforcement learning system in which an authorization engine adapts an ABAC model, as shown in Figure~\ref{fig:adaptive_rl_policy_learning}. The model depends on interacting with the administrator of the system to receive the feedback that helps the model make an authorization decision. The authors suggested four methods for initializing the learning model based on attribute value hierarchy to speed up the learning process. The paper focus on developing an adaptive ABAC policy learning model for a home IoT environment.

The authors experiment on multiple data sets, including synthesized and real ones, to evaluate the proposed method properly. The synthesized authorization records are produced based on ABAC policies: a set of ABAC policies with manually written policy rules and one with randomly generated policy rules. The real dataset is built from the records provided by Amazon~\cite{AmazonKaggle2013} (SL:~\ref{row:amazon-kaggle} of Table~\ref{tab:real-world-dataset}). Each access tuple in this dataset corresponds to an employee’s request for a resource and shows whether the access was permitted or not. The access log consists of the employees’ attribute values and the resources’ identifier.

Generally, traditional access control systems are imitated in the context of modern systems that are dynamic, complex, and large-scale, where it is difficult to maintain an accurate access control state in the system for a human administrator. As a potential solution to this problem, Nobi et al.~\cite{nobi2022toward} propose Deep Learning Based Access Control (DLBAC) utilizing advancements in deep neural networks.
The DLBAC works directly with the users and resources metadata and obviates the need for attribute and policy engineering. The output of the DLBAC model is a trained neural network, which takes user and resource metadata as input and makes access control decisions.

The authors create a prototype of DLBAC using ResNet~\cite{he2016deep} neural network. The authors simulate complex real-world scenarios for the performance evaluation and develop eight synthetic datasets. The datasets and related source code are publicly available\footnote{\url{https://github.com/dlbac/DlbacAlpha}}. The authors also experiment with their proposed method with two Amazon datasets~\cite{AmazonUCI2011, AmazonKaggle2013}. The empirical evaluation reveals that the DLBAC can make more accurate access control decisions and generalize better than traditional policy mining approaches and classical ML-based systems. The evaluation results demonstrate a trade-off while managing the over-provision (permitting unauthorized accesses) and under-provision (denying desired accesses). The proposed experiment suggests that traditional policy-based systems are highly inefficient in denying access, whereas DLBAC can properly balance under- and over-provision.

Prior researches~\cite{cappelletti2019quality} highlight that a neural network-based system, such as DLBAC, might not be feasible for access control due to the lack of decision interpretation. As DLBAC makes access decisions based on its underlying black-box neural network, it is crucial to understand the decisions in human terms. The authors propose Integrated Gradient and Knowledge Transferring based understanding methods to overcome this challenge of comprehending DLBAC decisions. The first method ranks user and resource metadata with respect to their influence on an access decision in question using an attribute score. The latter approach extracts a decision tree to understand a decision in the form of traditional rules. 
The authors also suggest that their proposed understanding techniques could be utilized in other aspects of access control, such as access modification.
\begin{table*}[t]
\centering
  \caption{Summarizing Machine Learning for Access Control Decision}
  \resizebox{\textwidth}{!}{%
  \begin{tabular}{cccccc}
    \hline
    \rowcolor{Gray}
    Reference
    & Application
    & \multicolumn{1}{p{4cm}}{\centering Problem Considered}
    & \multicolumn{1}{p{2cm}}{\centering Access Control Model}
    & \multicolumn{1}{p{2.2cm}}{\centering ML Approach}
    & \multicolumn{1}{p{2.2cm}}{\centering Dataset Type}\\\hline
    
    Chang et al. 2006~\cite{chang2006access}  & 
    Not specified & 
    \multicolumn{1}{p{4cm}}{\centering A novel access control model with time constraint} &
    \multicolumn{1}{p{2cm}}{\centering Time-constraint Access Control}  &
    \multicolumn{1}{p{2cm}}{\centering SVM} &
    \multicolumn{1}{p{2cm}}{\centering Syn}\\
    
    \rowcolor{LightGray}
    Outchakoucht et al. 2017~\cite{outchakoucht2017dynamic}  & 
    IoT & 
    \multicolumn{1}{p{4cm}}{\centering Blockchain based access control policy} &
    \multicolumn{1}{p{2cm}}{\centering Blockchain based Access Control}  &
    \multicolumn{1}{p{2.2cm}}{\centering Reinforcement Learning} &
    \multicolumn{1}{p{2.2cm}}{\centering No Evaluation}\\

    Cappelletti et al. 2019~\cite{cappelletti2019quality} & Not specified & \multicolumn{1}{p{4cm}}{\centering Inferring ABAC policies from access logs} & 
    \multicolumn{1}{p{2cm}}{\centering ABAC} & 
    \multicolumn{1}{p{2.2cm}}{\centering DT, RF, SVM, MLP} & 
    \multicolumn{1}{p{2.2cm}}{\centering RW (SL: \ref{row:amazon-uci}, \ref{row:amazon-kaggle}, \ref{row:incident})}\\
    
    \rowcolor{LightGray}
    Khilar et al. 2019~\cite{khilar2019trust}  & 
    Cloud Computing & 
    \multicolumn{1}{p{4cm}}{\centering Policy for cloud resources based on the access history and behaviour} &
    \multicolumn{1}{p{2cm}}{\centering Trust-Based Access Control}  &
    \multicolumn{1}{p{2.2cm}}{\centering RF, DT, SVM, Neural Network, etc.} &
    \multicolumn{1}{p{2.2cm}}{\centering Not Specified}\\
    
    Srivastava et al. 2020~\cite{srivastava2020machine}  & 
    \multicolumn{1}{p{3cm}}{\centering Defense, airport, and healthcare} & 
    \multicolumn{1}{p{4cm}}{\centering A novel access control framework to decide accesses based on the genuineness of the requester} &
    \multicolumn{1}{p{2cm}}{\centering Risk Adaptive Access Control (RAdAC)}  &
    \multicolumn{1}{p{2.2cm}}{\centering Neural Network, RF} &
    \multicolumn{1}{p{2.2cm}}{\centering Not Specified}\\
    
    \rowcolor{LightGray}
    Liu et al. 2021~\cite{liu2021efficient}  & 
    Big Data \& IoT & 
    \multicolumn{1}{p{4cm}}{\centering Improves the policy decision point (PDP) of the ABAC model} &
    ABAC &
    \multicolumn{1}{p{2.2cm}}{\centering RF} &
    \multicolumn{1}{p{2.2cm}}{\centering RW (SL: \ref{row:amazon-kaggle})}\\
    
    Karimi et al. 2021~\cite{karimi2021adaptive}  & 
    IoT & 
    \multicolumn{1}{p{4cm}}{\centering Adaptive ABAC policy learning} &
    ABAC &
    \multicolumn{1}{p{2.2cm}}{\centering Reinforcement Learning} &
    \multicolumn{1}{p{2.2cm}}{\centering Syn \& RW (SL: \ref{row:amazon-kaggle})}\\
    
    \rowcolor{LightGray}
    Nobi et al. 2022~\cite{nobi2022toward}  & 
    Not specified & 
    \multicolumn{1}{p{4cm}}{\centering A deep neural network based access control model} &
    DLBAC &
    \multicolumn{1}{p{2.2cm}}{\centering ResNet} &
    \multicolumn{1}{p{2.2cm}}{\centering Syn \& RW (SL: \ref{row:amazon-uci}, \ref{row:amazon-kaggle})}\\
    
    \hline
  \end{tabular}
  }
  \label{tab:access-decision}
\end{table*}

\section{Open Challenges and Future Research Directions}
\label{sec:open-challenges}


\subsection{Understanding Access Control Decisions}
In general, in complex situations where access control system has much overlapping, the non-symbolic ML models make better access control decisions than the traditional access control policies or symbolic ML models~\cite{cappelletti2019quality}. A neural network or any other non-symbolic method has an innate ability to learn subtle differences among users/resources and their relationships. However, to obtain such superior performance in these non-symbolic models, we have to compromise with \emph{understandability}--- that is, understanding the reason for a specific access decision.

Even though the use of ML models for access control decision is still in its inception, a lack of understandability could hinder the growth of this potential direction. This kind of understanding is straightforward for a written policy and not that complicated for the symbolic ML models. For example, the policy itself is written with humanly understandable logical rules, which is also applicable for the policies if an ML model generates such policies.
In the case of symbolic ML methods, one can easily dig down its underlying tree to extract logical rules and learn the underlying reasoning for a decision. In contrast, in non-symbolic ML model based systems, e.g., DLBAC~\cite{nobi2022toward} or Karimi et al.~\cite{karimi2021adaptive}, a `black box' function makes access control decision and is not in a human-understandable form which is crucial for security sensitive domains.

\begin{figure}[t]
	\includegraphics[width=0.4\linewidth]{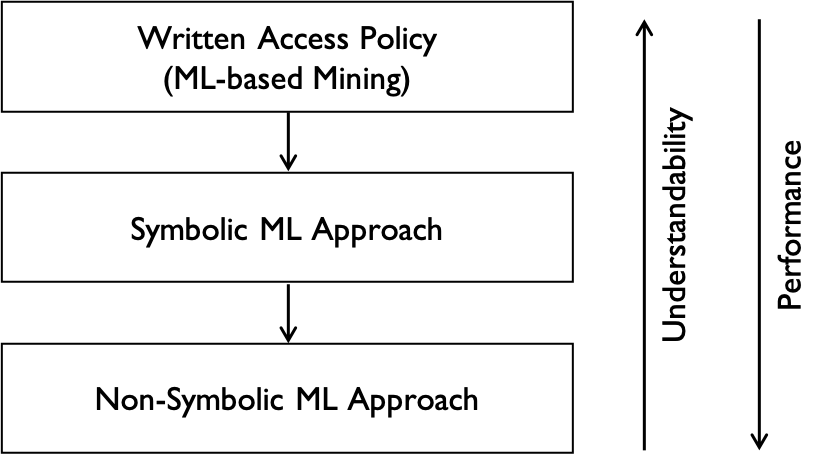}
	\caption{Trade-off Between Performance and Understandability in ML-based Systems.}
	\label{fig:understanding}
	\vspace{-3ex}
\end{figure}

As illustrated in Figure~\ref{fig:understanding}, as we go from the written policy to the non-symbolic approach, the performance goes up, though with the cost of a lack of understanding of decisions. This is one of the major limitations when choosing a non-symbolic ML model for access control. This issue is also applicable to other domains, including computer vision, malware analysis, financial systems, etc. The understanding issue, also known as interpretation or explainability in the typical ML arena, is a very active research area~\cite{islam2022systematic}. 
In many cases, it is observed that the solutions are domain-specific. For example, a \emph{computer vision} specific interpretation method would not directly work in the access control domain. 
Recently, Nobi et al.~\cite{nobi2022toward} explore this issue for the first time from access control perspectives and provide two methods for understanding neural network-based access control decisions in human terms to a large degree. 
However, as the proposed method does not guarantee the understanding of decisions with 100\% accuracy, it needs to study the area further to reach a better interpretation.

\subsection{Administration}
With the development of traditional access control or ML-based approaches, maintenance becomes crucial for regular access and policy management. In the access control domain, this problem is known as \textit{access control administration} and is crucial to keeping a system secured for the long run. In general, the administration is the task of accommodating the system's access authorization changes by modifying existing policy configurations or access control-related attributes. The administration's requirements could vary from model to model and system to system. In the case of RBAC, the administration could assign/ remove permission to the role, create a new role, manage role hierarchy, etc.~\cite{sandhu1999arbac99, bhatti2005x, singh2021role}. To accommodate changes in ABAC, one needs to adjust users/ resources attributes, create new rules, update existing rules, etc.~\cite{servos2017current, jha2016administrative}.
Similarly, ReBAC has relationship-related management and modifications to the policies to incorporate changes~\cite{stoller2015administrative, cheng2016extended}. Researchers thoroughly investigated administration issues for respective models in traditional access control models. 
As the application of machine learning for Access Control policy decisions is emerging, there are also related administration problems such as updating the corresponding ML model, adjusting access control-related information, etc. Currently, there is no approach to address administration issues from ML and access control perspectives. Therefore, it needs to explore the domain to formally define administration requirements for ML-based systems and find solutions for tackling underlying issues.

\subsection{Adversarial Attacks} The adversarial attack is a common concern for any ML-based system. An adversary can obtain unwarranted decisions~\cite{ballet2019imperceptible} by fooling the network with adversarial samples that are indistinguishable from natural ones by human~\cite{zheng2019distributionally}. It is an active research area in machine learning and studies this issue to design novel attacks that help determine loopholes in an ML model. There are several adversarial attacks available~\cite{kurakin2016adversarial, carlini2017towards}, though most of them are applicable for computer vision-oriented applications. Also, there are efforts to develop prevention methods to tackle these attacks while the model is in deployment. Adversarial training is one of the most successful defense methods~\cite{madry2017towards, hosseini2017blocking} where a model is trained with a combination of both natural and adversarial samples. 
With the application of ML in access control, this issue is also inevitable in the domain. In the context of access control, an adversarial attack could force an ML-assisted access control system to obtain unauthorized access. The attacker could \emph{trick} the system deliberately providing manipulated information of users and resources to prevail the desired access. Moreover, an ML-based access control system could be susceptible to the \emph{attribute-hiding attack} where an attacker may attempt to secure access by hiding (or removing) a portion of their information. Therefore, it is crucial to investigate adversarial attacks from access control perspectives and establish solutions to protect a system against such vulnerabilities.



\subsection{Lack of `Good' Datasets}
For access control-related research and development, it is necessary to have a ground truth to determine the efficacy of many applications, including role mining, attribute engineering, policy mining, etc. The performance of developed methods in these areas highly depends on the availability of high-quality real-world data~\cite{molloy2011adversaries}. This requirement is critical in the case of ML-based methods. Until now, only a few access control-related datasets are publicly available for research and experiments (Table~\ref{tab:real-world-dataset} reports publicly available real-world datasets). Also, the quality of these available datasets is not up to the mark as the datasets are highly imbalanced~\cite{cotrini2018mining, nobi2022toward} and contain less access control related information~\cite{heaps2021access}, etc. Therefore, it is not feasible to design an appropriate system based on those available datasets in many cases. For proper balancing of these available real-world datasets, often, the authors rely on extensive data preprocessing~\cite{liu2021efficient}, augmenting available datasets~\cite{alohaly2018deep, alohaly2019towards, zhou2019automatic}, amalgamating data from various sources~\cite{heaps2021access}, etc. Also, it is ubiquitous of generating synthetic access control datasets~\cite{xu2014mining} and applying them for different researches~\cite{argento2018towards, mocanu2015towards} to evaluate proposed approaches. In some cases, only real-world datasets or synthetically developed datasets are not sufficient due to the underlying issues in the datasets. To get rid of that, many researchers follow a hybrid approach by applying datasets from both publicly available real-world datasets and simulated datasets~\cite{bui2020learning, jabal2020polisma, karimi2021automatic, nobi2022toward}.

Even though we see the growing trend of the application of ML in access control, the lack of quality datasets from real-world organizations may be a critical blocking point. Many publicly available datasets are either highly anonymous or do not contain complete access control information. Besides, those real-world and simulated datasets often lack semantics and granularity of permissions~\cite{molloy2011adversaries} and are incapable of expressing an entire access control state of a system. Therefore, to advance in ML-based access control, the availability of a high-quality dataset is essential. 

\subsection{Bias and Fairness}
Various methods apply synthetic and real-world datasets to train respective machine learning models. The synthetic datasets are well crafted, considering different aspects of the data and access control characteristics in mind. Also, the datasets are designed to keep an equal portion of samples from all the classes (e.g., permit and deny are two classes while training a model for access decisions). However, this is not practical nor feasible for real-world datasets. For instance, the Amazon dataset~\cite{AmazonKaggle2013} holds historical access data of two consecutive years where employees were manually allowed or denied access to resources over time. The ratio between access grant and deny is highly imbalanced, where more than 90\% of the requests got access to the desired resources~\cite{cotrini2018mining, nobi2022toward}. In other words, the majority of the training data are from \emph{grant} class. Such disproportionate training data will force the trained model, irrespective of the type of machine learning model and architecture, to make \emph{biased} decision towards the grant class~\cite{nobi2022toward}. Therefore, to obtain a fair and reliable system, it is critical to understand the training data characteristics by auditing them before initiating the training phase~\cite{mehrabi2019survey}. Moreover, besides evaluating a system for access control decisions, it is equally important to evaluate the \emph{fairness} performance and establish a feedback loop~\cite{cutillo2020machine, nobi2022toward}.


\subsection{Insufficient Tools for Verification}
A reliable system design is usually ensured through a rigorous \emph{testing} and \emph{verification}. Testing is the evaluation of a system in several conditions to observe underlying behavior and detect errors. In contrast, the verification ensures that the system will not demonstrate any misbehavior under more general circumstances~\cite{goodfellow2017challenge}. Likewise, in access control, the implemented policy is verified and tested akin to verifying the correctness of software functionality (i.e., software testing)~\cite{hu2016ACPolicyVerification}. 
It is required to ensure that the engineered access control rules make correct access control decisions, which is complex and needs expensive effort. Unlike many domains, failing to verify the access control system's correctness to a large extent may have serious consequences, including but not limited to over-provision, under-provision, adversarial attacks, etc. However, this area is well studied for traditional access control systems, and there are methods for proper access control verification~\cite{kuhn2016testingKuhn, hu2021verificationNIST}. 

When the machine learning method is applied to any system, the system's performance is measured based on an unseen \emph{test dataset} to test the learned models' correctness to some degree. Evidently, this testing method cannot find all possible cases that may be misclassified. Also, such misclassification does not have an equal impact (or cost) on every application, and it could vary from domain to domain and application to application.
For example, granting an access request for an unauthorized resource could be more expensive than denying a legit request in an access control context. Therefore, a comprehensive verification is crucial before deploying a machine learning-assisted access control system. There are methods to verify a machine learning model automatically~\cite{leofante2018automated}. Also, each method has its own merits and demerits, and no one method will work for all the domains. Consequently, further research is needed to design a systematic verification and testing framework from machine learning and access control perspectives. 
\section{Summary}
\label{sec:conclusion}

This survey paper comprehensively explores the access control literature that uses machine learning approaches to solve various access control-related problems. We summarize that the benefits of machine learning are emerging in various areas of access control, including but not limited to attribute engineering, policy mining, access control policy verification, etc. We also discovered multiple efforts to utilize a trained machine learning model to decide access control requests as a substitute for a written access policy. In addition, we outline public real-world datasets used for machine learning-based access control research. We also propose a taxonomy of machine learning for access control and briefly discuss each approach following the proposed classification. Finally, we portray our observations regarding open challenges in the domain and provide potential guidelines to overcome them.



\bibliographystyle{ACM-Reference-Format}
\bibliography{references}


\end{document}